\documentclass[%
  reprint,
  superscriptaddress,
  longbibliography,
  preprintnumbers,
  amsmath,
  amssymb,
  aps,
  prl,
  floatfix,
  tightenlines
]{revtex4-2}

\usepackage{physics}
\usepackage{booktabs}
\usepackage{graphicx}
\usepackage{dcolumn}
\usepackage{bm}
\usepackage[percent]{overpic}

\usepackage{changes}
\usepackage[colorlinks=true,
citecolor=green,
linkcolor=purple,
anchorcolor=black,
urlcolor=purple]{hyperref}

\begin{document}

\title{A Qudit-Native Framework for Discrete Time Crystals}

\author{Wei-Guo Ma}
\affiliation{Beijing National Laboratory for Condensed Matter Physics,
Institute of Physics, Chinese Academy of Sciences, Beijing 100190, China}
\affiliation{School of Physical Sciences, University of Chinese Academy of Sciences, Beijing 100049, China}

\author{Heng Fan}
\email{hfan@iphy.ac.cn}
\affiliation{Beijing National Laboratory for Condensed Matter Physics,
Institute of Physics, Chinese Academy of Sciences, Beijing 100190, China}
\affiliation{School of Physical Sciences, University of Chinese Academy of Sciences, Beijing 100049, China}
\affiliation{Beijing Key Laboratory of Advanced Quantum Technology,
Beijing Academy of Quantum Information Sciences, Beijing 100193, China}
\affiliation{Hefei National Laboratory, Hefei 230088, China}
\affiliation{Songshan Lake Materials Laboratory, Dongguan, Guangdong 523808, China}

\author{Shi-Xin Zhang}
\email{shixinzhang@iphy.ac.cn}
\affiliation{Beijing National Laboratory for Condensed Matter Physics,
Institute of Physics, Chinese Academy of Sciences, Beijing 100190, China}

\date{\today}

\begin{abstract}
We introduce a qudit-native framework for engineering rich and robust discrete time crystals (DTCs) by leveraging their internal multilevel structure. Unlike in qubit systems, qudit-based DTCs exhibit distinct dynamical mechanisms that arise only in multilevel systems, as supported by a dressed normal-form analysis in the heating-suppression regime. These mechanisms are manifested in representative systems: we show that subspace-selective embedded kicks stabilize higher-order subharmonic responses and suppress thermalization, as demonstrated in spin-1 chains; in spin-3/2 systems, extending embedded kicks to more levels enables different level partitions and reveals that DTC robustness is dictated by the symmetry of the partition; and in spin-2 platforms, we realize concurrent 2T and 3T DTCs under a unified drive. These findings establish a systematic, hardware-efficient methodology for designing stable and multifunctional Floquet phases of matter on modern qudit-based quantum processors.
\end{abstract}

\maketitle
\textit{Introduction.}---Periodically driven many-body systems can host unique phases without equilibrium analogs. A discrete time crystal (DTC) is the canonical example where the system responds with robust subharmonic oscillations that spontaneously break the drive's discrete time-translation symmetry~\cite{DTC_review_Zaletel2023,annurev-conmatphys-031119-050658,FloquetTC_Else2016,PhysRevLett.116.250401,PhysRevB.93.245146,PhysRevB.94.085112, PhysRevB.99.195133}. Following extensive theoretical work ranging from realization mechanisms to applications~\cite{PhysRevA.91.033617,Clock_models_Yao2017,PhysRevB.95.214307,PhysRevLett.120.040404,PhysRevLett.120.110603,PhysRevLett.123.210602,PhysRevLett.125.060601,Yao:2020aa,PhysRevResearch.3.L042023,PhysRevLett.127.043602,Cabot2024PRL,Camacho2024PRR,Penner2025PRB,BarLev2024PRL,Saha2024PRA,PhysRevLett.126.020602,PhysRevA.105.013710}, signatures of time-crystalline behavior have been experimentally observed across a remarkable range of platforms~\cite{DTC_exp_trapped_ions_Zhang2017,DTC_exp_NV_Choi2017,DTC_exp_NMR_Rovny2018,MBL_DTC_Mi2022,Zhang:2022aa,Xiang2024NatComm,doi:10.1126/sciadv.abm7652,doi:10.1126/science.abk0603,Wu2024NatPhys,Wang:2025aa,PhysRevLett.132.183803,Bao:2024aa,1c1k-bv7z,xu2021realizingdiscretetimecrystal,zhang2025robustefficientquantumreservoir,camacho2024observingdlti}. The long-term stability of these phases hinges on the suppression of heating, a feat typically achieved through mechanisms such as many-body localization or high-frequency prethermal protection~\cite{Prethermal_DTC_Kyprianidis2021,Prethermal_DTC_Stasiuk2023,PhysRevLett.130.120403,PhysRevLett.115.256803,PhysRevLett.116.120401,KUWAHARA201696,Abanin:2017aa,PhysRevB.95.014112,PhysRevX.10.021046,PhysRevLett.127.050602,PhysRevX.4.031027, hl8q-4wy9}. Beyond the canonical period-doubling response, a wide phenomenology has also been established that includes higher-order and fractional temporal orders~\cite{Clock_models_Surace2019,HigherOrder_Fractional_Pizzi2021,HigherOrder_Rydberg_LiuNatComm2024,BarLev2024PRL,Penner2025PRB, PhysRevB.109.174310}, revealing a rich and diverse landscape of DTCs.

While the DTC paradigm is well-established for qubits, realizing it in higher-dimensional qudits presents unique challenges. This enlarged Hilbert space offers the potential for richer physics while also introducing greater complexity into the many-body dynamics. Prior DTC clock constructions~\cite{Clock_models_Yao2017,Clock_models_Surace2019,HigherOrder_Fractional_Pizzi2021,HigherOrder_Rydberg_LiuNatComm2024} realize the cycle with kick unitaries acting on all on-site levels, e.g., global $\pi$ rotations or homogeneous transverse fields. This approach suffices for two-level systems, where any DTC implementation necessarily addresses the full Hilbert space, but it becomes insufficient for higher-dimensional qudits. In a qudit, such global operations lack the selectivity to distinguish between internal levels, potentially moving population outside the intended active subspaces or failing to enforce specific subharmonic responses. To overcome this challenge, we must leverage the capabilities of modern platforms with native multilevel control, such as superconducting circuits, Rydberg arrays, solid-state spins and trapped ions~\cite{Qudit_SC_LiuPRX2023,vbh4-lysv,Qudit_SC_NguyenNatComm2024,PRXQuantum.4.030327,Goss:2022aa,PhysRevLett.134.050601,Qudit_Rydberg_EveredNature2023,HigherOrder_Rydberg_LiuNatComm2024,Qudit_SolidState_GuoPRL2024,quantumdot_qudit,Hrmo:2023aa,Ringbauer:2022aa}. We treat the multilevel Hilbert space as a design resource, implementing cycles confined to chosen on-site subspaces. This qudit-native construction suppresses unwanted leakage compared with global drives, and offers direct control of spectral sharpness and lifetime through on-site kick choices. Beyond selecting a single subspace, we can also compose disjoint on-site cycles with different periods to realize multiple subharmonic channels in parallel within one drive.

In this Letter, we use this framework to build qudit-based DTCs across different local dimensions $d$, focusing on $d=3, 4, 5$ as representative cases, and utilize a dressed normal-form analysis~\cite{PhysRevX.7.011026,PhysRevX.10.021046} to identify the mechanisms and stability underlying their subharmonic response. For spin-1 systems, we realize both period-doubled and period-tripled responses, and find that embedded subspace-selective kicks yield sharper Fourier peaks than global kicks by reducing population leakage into inactive levels. In spin-$3/2$ systems, we show that the stability of the phase depends on the symmetry of the chosen subspace partition; specifically, pairings symmetry-related levels outperform contiguous pairings. We further show that a spin-2 mixed trimer-doublet design supports concurrent $1/3$ and $1/2$ frequencies within a unified drive, with weights set by the initial state's support on each block. These findings provide a hardware-efficient, qudit-native framework for constructing stable qudit-based DTCs and enables concurrent subharmonic oscillations on multilevel platforms.

\begin{figure}[t]
\centering
\includegraphics[width=0.9\linewidth]{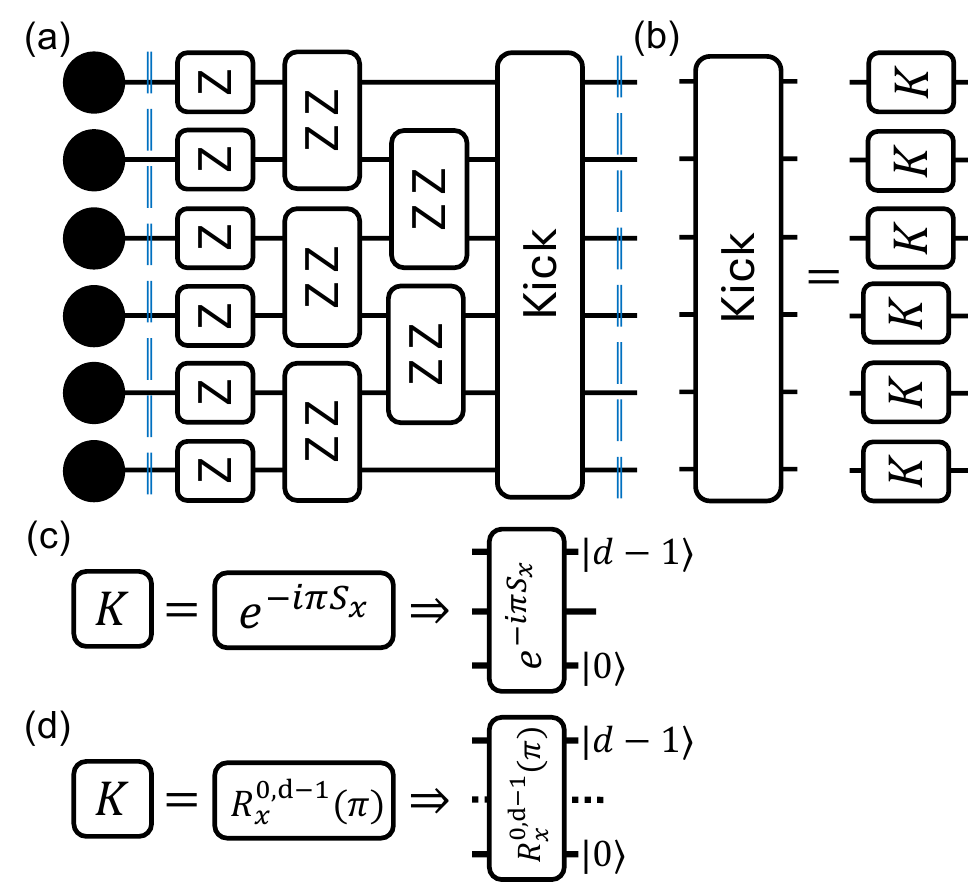}
\caption{(a) Qudit Floquet architecture. Each period factors as $Ke^{-iH_z}$: a diagonal phase layer generated by disordered on-site $Z_i$ fields and disordered nearest-neighbor $Z_i Z_{i+1}$ couplings, followed by a site-factorized kick $K$ that reshuffles populations.
(b) The Floquet kick factorizes into a tensor product of single-site operations. 
For a 2T DTC in a $d$-level system: (c) global kick via a single rotation with angle $\pi$ acting on the full on-site multiplet. (d) Embedded-kick realization confined to an active subspace (e.g., the doublet $\ket{0},\ket{d-1}$), leaving the complement inert.}
\label{fig:general_schematic}
\end{figure}

\textit{Setups and dressed normal form.}---We study DTC in $d$-level qudit chains within a heating-suppressed window~\cite{PhysRevLett.115.256803, PhysRevB.95.014112, PhysRevLett.116.120401, Saha2024PRA, KUWAHARA201696, Abanin:2017aa}, where the onset of infinite-temperature chaos is greatly delayed and suppressed. We introduce the static layer as a disordered Ising Hamiltonian~\cite{PhysRevLett.111.127201, PhysRevB.90.174202, PhysRevLett.95.206603, BASKO20061126,PhysRevB.75.155111,PhysRevB.77.064426,PhysRevB.81.134202,Cuevas:2012aa,PhysRevLett.109.017202,PhysRevLett.110.067204,PhysRevLett.110.260601},
\begin{equation}
H_z=\sum_{i=0}^{N-2} J^{z}_{i,i+1}\,S_i^z S_{i+1}^z+\sum_{i=0}^{N-1} h_i\, S_i^z,
\label{eq:Hz}
\end{equation}
where $h_i\in[-W_h,W_h]$ and $J^{z}_{i,i+1}\in[J^z{-}W_J,J^z{+}W_J]$ are random fields and couplings under open boundary conditions. We adopt $J^z{=}1$ as the characteristic energy scale and set the disorder strengths to $W_h{=}6$ and $W_J{=}1$ for all numerical simulations, unless otherwise specified. These terms commute and, thus, $e^{-iH_z}$ adds only configuration-dependent phases. As illustrated in Fig.~\!\ref{fig:general_schematic}(a), the periodic dynamics are formulated as a quantum circuit model. In this representation, the $S^z{=}\tfrac12\mathrm{diag}(-d{+}1,-d{+}3,\dots,d{-}3,d{-}1)$ operator is realized by the generalized qudit gate $Z_d{=}\exp\!\big[i\tfrac{2\pi}{d}(S^z{+}\tfrac{d-1}{2})\big]$. Each Floquet period is $U_F(\varepsilon)=K_m(\varepsilon)e^{-iH_z}$ with kick $K_m(\varepsilon)=\bigotimes_{i=0}^{N-1}K_{m, i}(\varepsilon)$ acting within each local $d$-level multiplet, and $\varepsilon$ quantifies kick imperfections. 

We model the imperfections as multiplicative angle errors: for a target on-site $m$-cycle with angle $\theta_0$ and generator $G$, we implement angle imperfection $\theta_0{\to}(1{+}\varepsilon)\theta_0$ with the same relative error $\varepsilon$ for every application of the on-site gate, and write $K_m(\varepsilon){=}K_{m}E(\varepsilon)$, where $K_{m}$ is the ideal $m$-cycle driver with $K_{m}^{m}{=}I$ and $E(\varepsilon){=}\exp[-i\varepsilon G_1{+}\mathcal{O}(\varepsilon^2)]$ collects control errors with $G_1=\theta_0 G$ (see~\cite{supplementary}). Any operator admits a time-charge decomposition as $X{=}\sum_q X_q$ with $K_{m} X_q K_{m}^\dagger{=}e^{2\pi i q/m}X_q$ (neutral $q{=}0$, charged $q{\neq}0$). In the heating-suppressed window, we can derive the Floquet operator as a dressed normal form as~\cite{PhysRevX.7.011026,PhysRevX.10.021046}
\begin{equation}
V(\varepsilon)U_F(\varepsilon)V^\dagger(\varepsilon){=}K_{m}\,e^{-iD(\varepsilon)}e^{-iR(\varepsilon)},
  \label{eq:normalform}
\end{equation}
where $V(\varepsilon)$ is near identity and $[K_{m},D(\varepsilon)]=0$. The neutral part is  $D(\varepsilon)=D_0+\delta D(\varepsilon)$ with $D_0=\tfrac{1}{m}\sum_{j=0}^{m-1}K_{m}^{-j}H_z K_{m}^{j}$ and $\delta D(\varepsilon)=\tfrac{\varepsilon}{2}[D_0,G_{1,0}]+\mathcal{O}(\varepsilon^2)$. The charged remainder is $R(\varepsilon)=\varepsilon\sum_{q\neq 0}\!\big(G_{1,q}+\tfrac{i}{2}[D_0,G_{1,q}]\big)+\mathcal{O}(\varepsilon^2)$.

The dressed form of an observable $O$ is defined as $\widetilde O{=}V(\varepsilon)OV^\dagger(\varepsilon)$. We project $\widetilde O$ into the $\mathbb{Z}_m$ time-charge sectors, with components $O_q=\tfrac{1}{m}\sum_{j=0}^{m-1} e^{-2\pi i q j/m}K_m^{j}\widetilde O K_m^{-j}$. Let $\ket{\widetilde\psi_0}{=}V(\varepsilon)\ket{\psi_0}$ denotes the dressed state, and the $n$-period stroboscopic expectation is
\begin{align}
&\expval{O(n)} = \sum_{q=0}^{m-1}\expval{e^{inD}K_m^{-n}O_qK_m^{n}e^{-inD}}{\widetilde\psi_0} + \mathcal{O}(R(\varepsilon)) \nonumber \\
&= \sum_{q=0}^{m-1}e^{i2\pi q n/m}\expval{e^{inD(\varepsilon)}O_q e^{-inD(\varepsilon)}}{\widetilde\psi_0} + \mathcal{O}(R(\varepsilon)).
\label{eq:On}
\end{align}

Eq.~\!\eqref{eq:On} reveals distinct roles for the effective terms. Kick $K_m$ locks the subharmonic frequency at $f_q = q/m$. Since $[K_m, D]=0$, the neutral term $D$ preserves the $q$-sectors, causing only pure dephasing of the response envelope without shifting the central frequency. In contrast, the charged term $R$ mixes different $q$-sectors. This mixing scatters spectral weight away from the locked subharmonic frequencies, effectively melting the order parameter. For our specific imperfection model where $G_1$ commutes with $K_m$, the charged terms vanish in linear order ($R \sim \varepsilon^2$), ensuring that the locked frequency shift is suppressed.

Numerically, we monitor the chain-averaged magnetization~\cite{Prethermal_DTC_Kyprianidis2021,MBL_DTC_Mi2022}
\begin{equation}
  M_z(n)=\frac{1}{N}\sum_i \langle S_i^z(nT)\rangle.
  \label{eq:magnetization}
\end{equation}
We quantify the subharmonic response by the subharmonic weight~\cite{PhysRevLett.122.043603,PhysRevLett.127.090602}, defined as the normalized Fourier coefficient of the chain-averaged magnetization at target frequency $1/m$
\begin{equation}
C_m=\frac{S(k_m)}{\sum_{k=0}^{N_t-1}S(k)},\,
S(k)=\left|\sum_{n=0}^{N_t-1}\big[M_z(n)-\overline{M}_z\big] e^{-i\frac{2\pi kn}{N_t}}\right|^2,
\end{equation}
where $\overline{M}_z=(1/N_t)\sum_{n}M_z(n)$ is the time average, $N_t=300$ is the number of recorded periods, and $k_m$ corresponds to frequency $1/m$. $C_m$ thus measures the spectral weight at the subharmonic frequency $1/m$ and provides a compact indicator of the melting of the time-crystalline response. All the simulations are carried out using {\sf TensorCircuit-NG}~\cite{Zhang2022Tc, zhang2026tensorcircuit}.

\begin{figure}[t]
\centering
\includegraphics[width=\linewidth]{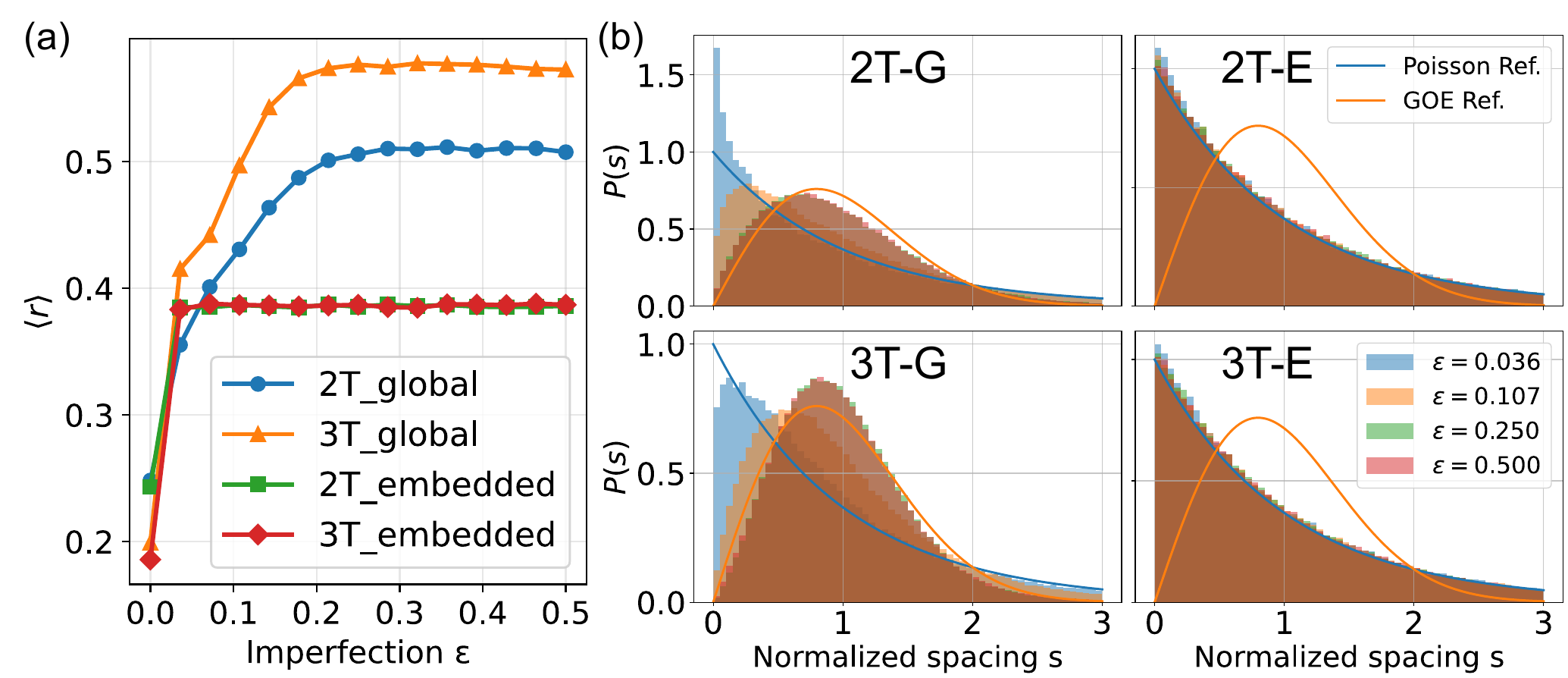}
\caption{Resistance to thermalization in spin-1 chains ($N=8$). 
(a) Average adjacent-gap ratio $\langle r \rangle$ versus drive imperfection $\varepsilon$. Embedded 2T and 3T drives (green and red lines) remain pinned to the Poisson value ($\approx 0.39$), indicating that the effective Hamiltonian remains localized or integrable-like. In contrast, global drives (blue and orange lines) flow toward the GOE limit ($\approx 0.53$), marking the onset of many-body chaos.
(b) Level spacing distributions $P(s)$. While global drives (left) develop Wigner-Dyson level repulsion at high $\varepsilon$, embedded drives (right) retain Poissonian statistics across the full range of imperfections, confirming that subspace selectivity effectively suppresses sector mixing.}
\label{fig:r_vs_eps}
\end{figure}

\textit{Spin-1: diagnostics and results.}---We instantiate the framework in spin-1 chains ($d{=}3$) by realizing period-doubled (2T) and period-tripled (3T) phases under the disordered static layer Eq.~\!\eqref{eq:Hz}. As a natural generalization of standard qubit protocols, as shown in Fig.~\!\ref{fig:general_schematic}(c), the global protocols apply system-wide rotations $K^{\mathrm{glob}}_m = \exp[-i\frac{2\pi}{m}(1+\varepsilon)S^x]$. In contrast, motivated by the multi-level structure, we introduce embedded kicks where the drive is confined to a specific subspace, as illustrated in Fig.~\!\ref{fig:general_schematic}(d). For examples, the embedded 2T drive flips the $\{\ket{0}, \ket{2}\}$ subspace via selective pulses while leaving $\ket{1}$ idle, and the embedded 3T drive implements an effective permutation cycle with two embedded kicks: $\ket{0}{\to}\ket{2}{\to}\ket{1}{\to}\ket{0}$ (see gate sequences in~\cite{supplementary}). Within the normal form, such global addressing diffuses the probability density across the entire multiplet, undermining the heating-suppressed regime and melting the time-crystalline order.

\begin{figure}[t]
\centering
\includegraphics[width=\linewidth]{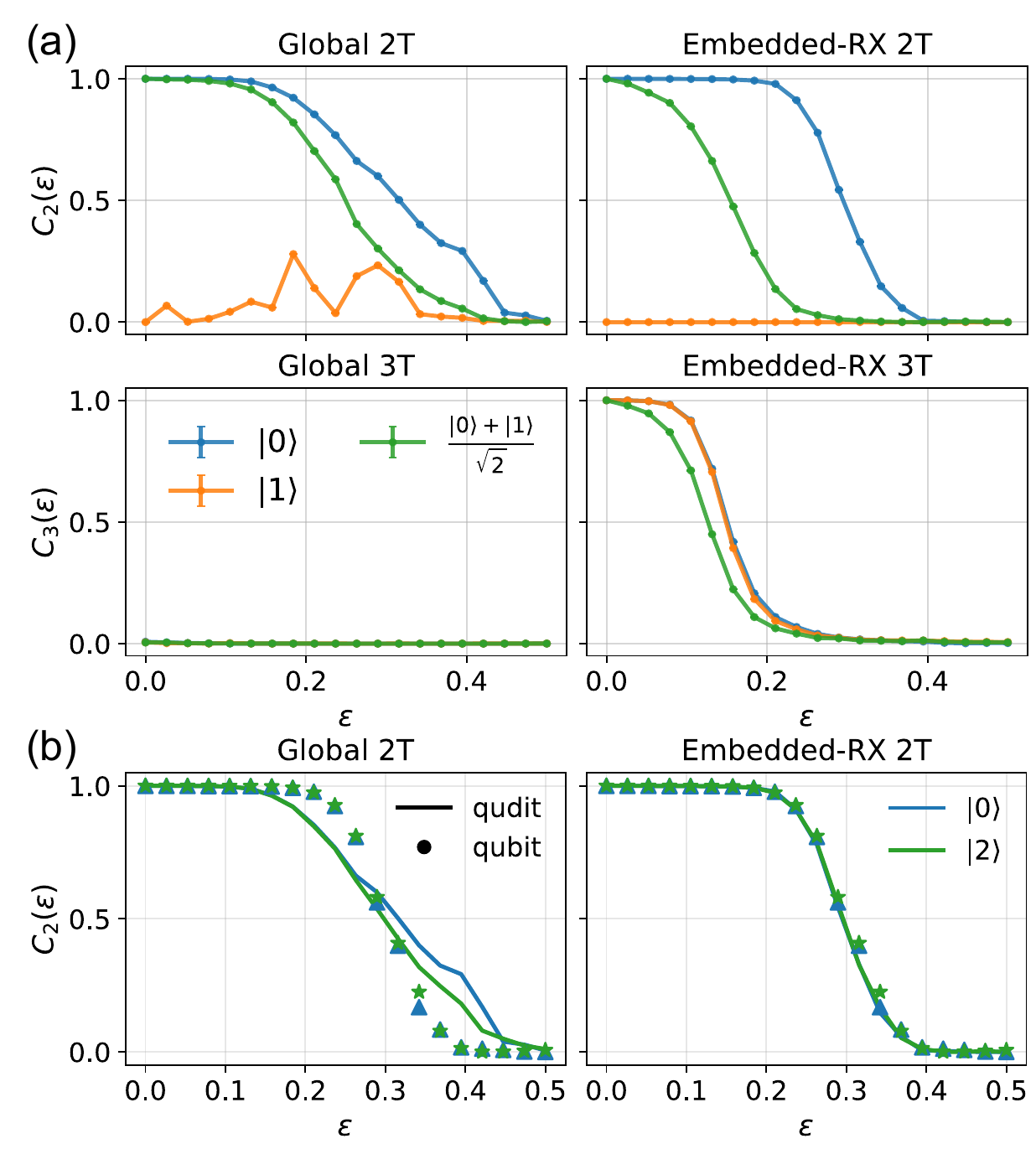}
\caption{Robustness of spin-1 DTCs ($N{=}14$). 
(a) Subharmonic weight $C_m$ vs.~imperfection $\varepsilon$. The embedded 2T protocol yields a broad plateau, whereas the global 2T decays rapidly due to leakage into the idle $\ket{1}$ state. For 3T, the embedded produces a stable response, while the global fails to maintain order due to basis mixing. Different colors represent various initial state preparations.
(b) Comparison with a qubit baseline. 
The embedded 2T data (solid lines) tracks the response of an ideal qubit system (symbols) mapped to the $\{\ket{0}, \ket{2}\}$ subspace, confirming that the protocol successfully hides the extra dimension. The global drive deviates significantly, highlighting the detrimental cost of global addressing.}
\label{fig:D3}
\end{figure}

We first characterize the spectral statistics of the Floquet operator $U_F$ within fixed symmetry sectors to confirm that the dynamics lies in a heating-suppressed regime where the dressed normal form in Eq.~\!\eqref{eq:normalform} applies. As a compact scalar diagnostic we use the average adjacent-gap ratio $r$~\cite{PhysRevB.75.155111}, defined for consecutive quasi-energy spacings. $r\approx 0.386$ indicates Poisson (localized) statistics, while $r\approx 0.53$ indicates Gaussian orthogonal ensemble (GOE) thermalizing statistics. Fig.~\!\ref{fig:r_vs_eps} shows $r(\varepsilon)$ for chain length $N=8$ together with spacing histograms. Embedded protocols remain near the Poisson value throughout the imperfection window used for our benchmarks. In contrast, global protocols drift toward the GOE value as $\varepsilon$ increases, signaling thermalization with the practical consequence that subharmonic features melt: the locked frequency disappears and the corresponding spectral weight decays.

Fig.~\!\ref{fig:D3}(a) presents the subharmonic weight $C_m(\varepsilon)$ for $N=14$. For the 2T phase, the embedded protocol exhibits a robust plateau up to larger $\varepsilon$, significantly outperforming the global drive. To isolate the origin of this stability, we compare against a spin-$1/2$ baseline (Fig.~\!\ref{fig:D3}(b)) obtained by mapping the active $\{\ket{0}, \ket{2}\}$ doublet to a qubit. The embedded 2T result tracks this qubit baseline perfectly, demonstrating that our protocol successfully eliminates the leakage of the larger Hilbert space. The global 2T drive, lacking this isolation, suffers from enhanced neutral-sector dephasing. The contrast is even sharper for the 3T case. The embedded cycle yields a clear $1/3$ response by preserving the block-diagonal structure of the effective Hamiltonian. However, the global 3T drive fails to lock even at weak imperfection; the global rotation inherently mixes the $S^z$ with $S^y$, suppressing the projection of the subharmonic signal onto the observable. We provide detailed derivations and an analysis across the parameter regimes of Eq.~\!\eqref{eq:Hz} in the supplemental material~\cite{supplementary}, further corroborating the advancements of the embedded kicks.

\begin{figure}[t]
\centering
\includegraphics[width=\linewidth]{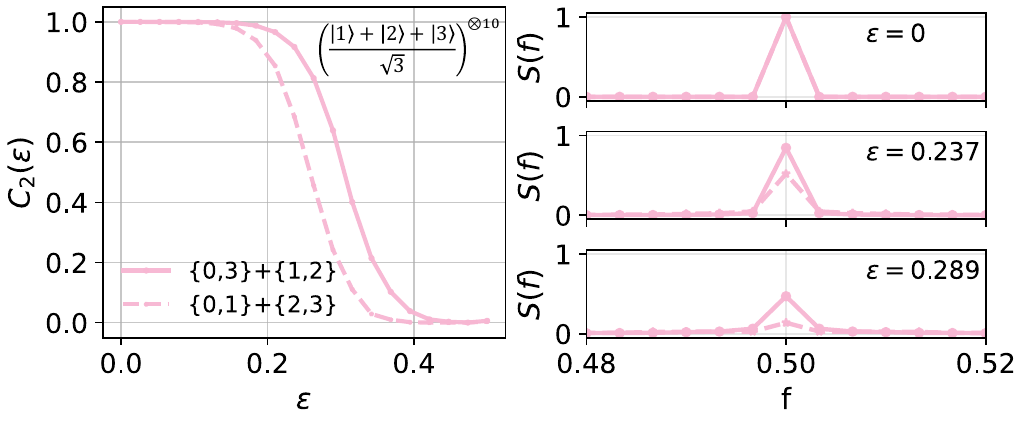}
\caption{Stability of doublet-partitioned spin-3/2 systems ($N{=}10$).
(Left) symmetric partition $\{0,3\}\oplus\{1,2\}$: broad near-unity $C_2$ plateau. Contiguous partition $\{0,1\}\oplus\{2,3\}$: less stability and the subharmonic weight drops for smaller $\varepsilon$. 
(Right) corresponding Fourier spectra at various $\varepsilon$, the contiguous partition exhibits significant broadening and lower peak with increasing imperfection.}
\label{fig:D4}
\end{figure}

\textit{Spin-3/2: level partitioning and symmetry.}---Building on the spin-1 result, the embedded kick can partition qudit manifolds into multiple independent sectors. This multi-sector capability unavailable to qubits, where the Hilbert space permits only a doublet. In spin-3/2 chains, the larger Hilbert space allows us to embed independent 2T cycles within disjoint subspaces, illustrating a tactical choice of subspace configuration.

Fig.~\!\ref{fig:D4} contrasts two such partitioning strategies ($N{=}10$): a symmetric partition pairing states with opposite magnetization ($\{0,3\}\oplus\{1,2\}$), and a contiguous partition pairing neighbor levels ($\{0,1\}\oplus\{2, 3\}$). The results reveal an interesting stability gap. The symmetric partition yields a broad, near-unity plateau in $C_2(\varepsilon)$. In contrast, the contiguous partition degrades significantly with smaller imperfection. This performance difference is explained by the structure of the neutral effective Hamiltonian $D_0 \approx \frac{1}{2}(H_z + K^\dagger H_z K)$. 

Note that the on-site fields average within each active doublet $b \in\{A,B\}$ as $\tfrac{1}{2}\left(S^z{+}P_2^{-1}S^z P_2\right){=}\mu_A\Pi_A{+}\mu_B\Pi_B$, where $\Pi_{A,B}$ project to the doublets and $\mu_{A,B}$ are block averages of $S^z$. For $\{0,3\}\oplus\{1,2\}$, where $(\mu_A,\mu_B)=(0,0)$, so on-site fields cancel in $D_0$ and block populations are preserved. For $\{0,1\}\oplus\{2,3\}$, $(\mu_A,\mu_B)=(-1,+1)$, yielding block-dependent neutral phases that accelerate the disorder-induced envelope dephasing. Coupling terms behave analogously and add block-asymmetric neutral contributions in the contiguous case. A design principle thus follows: choose the partition so that each active block has a vanishing mean of the operators entering $H_z$, which minimizes neutral dephasing contribution in $D_0$. In our minimal implementations, $G_1$ generates the same $\pi$ rotation as $P_2$ within each active doublet and is neutral under $\mathbb{Z}_2$, hence $R(\varepsilon)=\mathcal{O}(\varepsilon^2)$ for both partitions. The broader line and shorter plateau with the contiguous split therefore originate from stronger block-asymmetric neutral terms in $D_0$, instead of the charged residual $R(\varepsilon)$.

\begin{figure}[t]
\centering
\includegraphics[width=\linewidth]{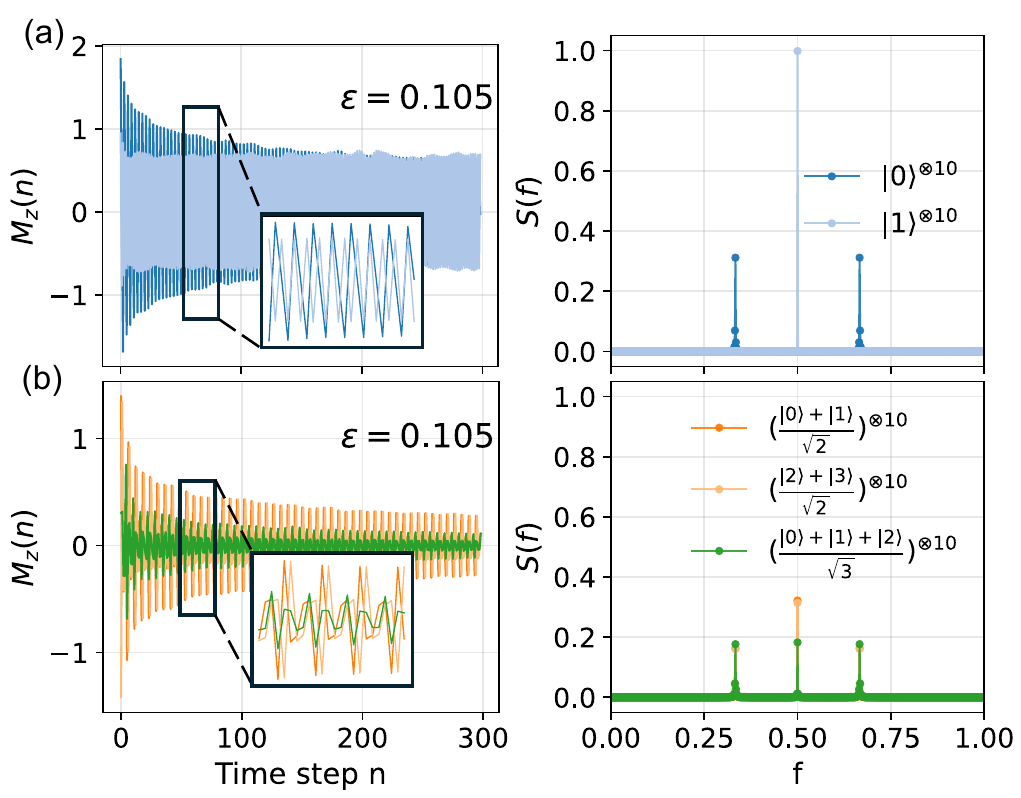}
\caption{Mixed trimer-doublet protocol on $N=10$ and $d=5$ qudits. Dynamics and Fourier spectra of $M_z$ show (a) single-line locking at $1/3$ (for trimer-only preparations) or $1/2$ (for doublet-only preparations), and (b) concurrent $1/3$ and $1/2$ components for initial states of cross-block superpositions, with relative weights determined by the initial support. Different colors represent various initial state preparations.}
\label{fig:d5_mixed}
\end{figure}

\textit{Spin-2: frequency-multiplexed DTCs.}---Finally, we demonstrate the capacity for parallel information processing within the multilevel structure of a single spatial site. We implement a mixed trimer-doublet protocol on an $N{=}10$, $d{=}5$ chain by partitioning each qudit with symmetric partition $\{0,2,4\}\oplus\{1,3\}$ (the embedded kick compilation is given in~\cite{supplementary}). A Floquet period produces a block-diagonal carrier
\begin{equation}
V(\varepsilon)U_F(\varepsilon)V^\dagger(\varepsilon)=(P_3\oplus P_2)e^{-i\left[D_0+\delta D(\varepsilon)\right]}e^{-iR(\varepsilon)},
\label{eq:d5_normalform}
\end{equation}
with $P_3^3=P_2^2=I$. The dressed observables thus decompose additively across the two blocks. When the initial state has support in both blocks, the magnetization contains coexisting $1/3$ and $1/2$ subharmonic components
\begin{equation}
\langle M_z(n)\rangle \approx 
E_{3}(n)e^{i\frac{2\pi}{3} n}
+ E_{2}(n)e^{i\pi n}
+ \text{c.c.} + \mathcal{O}(\varepsilon^2),
\label{eq:d5_mz_mix}
\end{equation}
where $E_{3}(n)$ and $E_{2}(n)$ are slowly varying block-resolved envelopes set by the corresponding neutral operators $D^{(\mathcal T)}(\varepsilon)$ and $D^{(\mathcal B)}(\varepsilon)$. As shown in Fig.~\!\ref{fig:d5_mixed} for $N=10$ qudits, the spectral response displays simultaneous peaks with relative weights linearly mapped from the initial population distribution. The longevity of this multiplexed signal is further corroborated by open-system results, which show that the subharmonic signatures endure under practical decoherence noise~\cite{supplementary}. This realization establishes the qudit not merely as a larger spin, but as a frequency-multiplexed unit capable of hosting complex quantum transformations.

\textit{Discussion and conclusion}.---We have established a qudit-native framework that leverages the multi-level space to construct rich and robust DTCs. In a heating-suppressed window, we identify the mechanisms governing stability through a dressed normal form, whereby the carrier $K_m$ locks the subharmonic frequency; the neutral term $D$ governs the envelope's pure dephasing; and the charged term $R$ drives the mixing of time-charge sectors resulting in frequency shifts. For the identical on-site imperfections considered in our study, the linear charged sector cancels, thus the differences in robustness are governed primarily by the neutral generator $D(\varepsilon)$. Within this regime, three neutral-sector design rules emerge: (i) confining the kick to an active on-site subspace restricts $D(\varepsilon)$ and mitigates dephasing; (ii) employing symmetric partitions of the local Hilbert space to enhance dynamical averaging in $D_0$ and partially cancel random fields and couplings; and (iii) utilizing a block-diagonal carrier, where each block admits its own $\mathbb Z_{m_s}$ oscillation dynamics, so multiple subharmonic channels at $1/m_s$ can coexist, with relative weights set by the initial state support.

Taken together, our framework provides a practical recipe for realizing stable, programmable, and multifunctional non-equilibrium phases of matter in multilevel platforms, well suited to hardware-efficient implementation on near-term quantum devices~\cite{Preskill2018quantumcomputingin, Arute:2019aa}. Although higher-level states in qudit platforms typically suffer from accelerated decoherence, recent experimental demonstrations of high-fidelity control~\cite{Qudit_SC_LiuPRX2023, vbh4-lysv} indicate that exploring rich physical phenomena in these systems is already feasible. Grounded in the experimental parameters reported in these platforms, our open-system analysis~\cite{supplementary} confirms that qudit-based DTC signatures can persist over experimentally relevant timescales, even under realistic thermal relaxation noise. While practical implementations may still be subject to many hardware constraints such as intra-period control conflicts and beyond-nearest-neighbor stray couplings, our results establish a viable foundation for exploring diverse DTCs in physical qudit platforms.

\textit{Acknowledgments}---We thank Zhen-Dong Cao for helpful discussions. This work was supported by Quantum Science and Technology-National Science and Technology Major Project (2024ZD0301700), the National Natural Science Foundation of China (Grants No. 92265207, No. T2121001, No. U25A6009, No. T2322030, No. 12122504, No. 12274142, and No. 12475017), QNMP (Grant No. 2021ZD0301800). 

\providecommand{\noopsort}[1]{}\providecommand{\singleletter}[1]{#1}%

\end{document}


\renewcommand{\thefigure}{S\arabic{figure}}
\setcounter{figure}{0}
\renewcommand{\theequation}{S\arabic{equation}} 
\setcounter{equation}{0}

\title{Supplemental Material for\\ ``A Qudit-Native Framework for Discrete Time Crystals''}

\author{Wei-Guo Ma}
\affiliation{Beijing National Laboratory for Condensed Matter Physics,
Institute of Physics, Chinese Academy of Sciences, Beijing 100190, China}
\affiliation{School of Physical Sciences, University of Chinese Academy of Sciences, Beijing 100049, China}

\author{Heng Fan}
\email{hfan@iphy.ac.cn}
\affiliation{Beijing National Laboratory for Condensed Matter Physics,
Institute of Physics, Chinese Academy of Sciences, Beijing 100190, China}
\affiliation{School of Physical Sciences, University of Chinese Academy of Sciences, Beijing 100049, China}
\affiliation{Beijing Key Laboratory of Advanced Quantum Technology,
Beijing Academy of Quantum Information Sciences, Beijing 100193, China}
\affiliation{Hefei National Laboratory, Hefei 230088, China}
\affiliation{Songshan Lake Materials Laboratory, Dongguan, Guangdong 523808, China}

\author{Shi-Xin Zhang}
\email{shixinzhang@iphy.ac.cn}
\affiliation{Beijing National Laboratory for Condensed Matter Physics,
Institute of Physics, Chinese Academy of Sciences, Beijing 100190, China}

\date{\today}

\maketitle
\section{A. Model, conventions, and notation}
\label{sec:A}
\subsection{A1. Static layer and disorder}
\label{sec:A1}
We consider a one-dimensional chain of $N$ qudits with $d$-dimensional Hilbert space. The static layer is
\begin{equation}
H_z=\sum_{i=0}^{N-2} J^{z}_{i,i+1} S_i^z S_{i+1}^z+\sum_{i=0}^{N-1} h_i S_i^z,
\label{eq:S1_Hz}
\end{equation}
where the fields and couplings (with disorder) are drawn independently from bounded intervals,
\begin{equation}
h_i\in[-W_h,W_h],\quad J^{z}_{i,i+1}\in[J^z-W_J, J^z+W_J].
\label{eq:S1_disorder}
\end{equation}
All terms in $H_z$ commute, so $e^{-iH_z}$ is stroboscopically exact and contributes only configuration-dependent phases. Unless stated otherwise, we assume open boundary conditions and adopt $J^z{=}1$ as the characteristic energy scale and set the disorder strengths to $W_h{=}6$ and $W_J{=}1$ for all numerical simulations, unless otherwise specified. On each site we fix the computational basis $\{\ket{m}\}_{m=0}^{d-1}$ and take
\begin{equation}
S^z=\mathrm{diag}\!\left(-\frac{d-1}{2},-\frac{d-3}{2},\dots,\frac{d-3}{2},\frac{d-1}{2}\right),
\label{eq:S1_Sz}
\end{equation}
so that $S^z\ket{m}=m_z \ket{m}$ with $m_z= m-\frac12(d-1)$. For $d\in\{3,4,5\}$ this gives
\begin{align}
d=3:&\;   m_z\in\{-1,0,1\},\nonumber\\
d=4:&\;   m_z\in\{-\frac{3}{2},-\frac{1}{2},\frac{1}{2},\frac{3}{2}\},\nonumber\\
d=5:&\;   m_z\in\{-2,-1,0,1,2\}. \nonumber
\end{align}
In circuit-level implementations the corresponding diagonal phase is realized by the generalized qudit gate 
\begin{equation}
  Z_d=\exp\left[i\frac{2\pi}{d}(S^z+\frac{d-1}{2})\right],
\end{equation}
which shares the eigenstructure of $S^z$ up to a global phase and is used when compiling purely diagonal rotations. We use $S^{x,y}$ to denote the standard Hermitian generators acting within the local $d$-level representation.

\subsection{A2. Floquet period and control-error model}
\label{sec:A2}
A single Floquet period factors as
\begin{equation}
U_F(\varepsilon)=K_m(\varepsilon) e^{-iH_z},
\label{eq:S1_UF}
\end{equation}
with a site-factorized kick $K_m(\varepsilon)=\bigotimes_{i=0}^{N-1}K_{m, i}(\varepsilon)$ that reshuffles populations only within specified on-site subspaces (``active blocks''). With an angle $\theta_0$ and a generator $G$, we write
\begin{equation}
K_m(\varepsilon)=K_m E(\varepsilon),\quad 
E(\varepsilon)=\exp\left[-i\varepsilon G_1+\mathcal{O}(\varepsilon^2)\right],
\label{eq:S1_Kerr}
\end{equation}
where $K_m$ is the ideal $m$-cycle carrier, and Eq.~\!\eqref{eq:S1_Kerr} defines a multiplicative angle-error model with small, dimensionless imperfection $\varepsilon$. Here $K_m$ is the ideal kick, $G_1=\theta_0G$ is the first-order error generator, and $\mathcal{O}(\varepsilon^2)$ collects second-order corrections that arise from noncommuting pieces within compiled embedded sequences.

\subsection{A3. Time-charge grading and notation}
\label{sec:A3}
Given an $m$-cycle ideal kick $K_m$ with $K_m^{m}=I$ acting locally on each site ($K_m=\otimes_{i=0}^{N-1}K_{m, i}$, e.g., a doublet swap for $m{=}2$ or a trimer cycle for $m{=}3$), we use the $\mathbb{Z}_m$ grading induced by $K_m$. Concretely, $K_m$ acts as the $m$-cycle on the chosen active subspace and as the identity on its inactive complement. The $\mathbb{Z}_m$ grading defined below is thus induced directly by $K_m$ and is invariant under neutral dressings that commute with $K_m$.

Any operator $X$ admits a decomposition
\begin{equation}
X=\sum_{q=0}^{m-1}X_q,\quad 
X_q=\frac{1}{m}\sum_{j=0}^{m-1}e^{-2\pi i q j/m}\,K_m^{\,j}\,X\,K_m^{-j},
\label{eq:S1_proj}
\end{equation}
so that
\begin{equation}
K_m\,X_q\,K_m^\dagger=e^{2\pi i q/m}X_q.
\label{eq:S1_chargecomm}
\end{equation}
We apply this convention uniformly to error generators and observables. In particular, we write $G_{1,q}$ for the $q$-charged component of $G_1$ and use the same subscript convention for all graded objects (e.g., $O_q$). Throughout, charges are understood modulo $m$, boldface indices (e.g., $i,i{+}1$) denote sites, and $q\in\{0,1,\dots,m-1\}$ labels time-charge sectors. For tensor-product operators, charges add under multiplication modulo $m$ with respect to the site-wise action of $K_m$.

\section{B. Dressed normal form and error contributions}
\label{sec:B}

\subsection{B1. Toggling-frame expansion}
\label{sec:B1}
We factor each Floquet period as
\begin{equation}
U_F(\varepsilon)=K_m(\varepsilon)e^{-iH_z},\quad
K_m(\varepsilon)=K_mE(\varepsilon),\quad
E(\varepsilon)=\exp\left[-i\varepsilon G_1+\mathcal{O}(\varepsilon^2)\right].
\label{eq:S2_Kerr}
\end{equation}
Passing to the toggling frame defined by $K_m$ gives
\begin{equation}
K_m^\dagger U_F(\varepsilon)=E(\varepsilon)e^{-iH_z}
=\exp\left\{-i\left(H_z+\varepsilon G_1\right)-\frac{\varepsilon}{2}\,[G_1,H_z]+\mathcal{O}(\varepsilon^2)\right\},
\label{eq:S2_toggling}
\end{equation}
where the Baker-Campbell-Hausdorff (BCH)/Magnus extension~\cite{BLANES2009151, RevModPhys.89.011004, KUWAHARA201696} is used to order $\varepsilon^2$. This representation will be used below to separate neutral pieces (commuting with $K_m$) and charged pieces (nonzero time charge).

\subsection{B2. Normal form and commuting structure}
\label{sec:B2}

In a heating-suppressed window~\cite{PhysRevB.95.014112,PhysRevLett.116.120401,Saha2024PRA}, there exists a quasi-local dressing $V(\varepsilon)$~\cite{PhysRevX.7.011026,PhysRevX.10.021046} such that
\begin{equation}
V(\varepsilon)\,U_F(\varepsilon)\,V^\dagger(\varepsilon)
=K_m e^{-i\left[D_0+\delta D(\varepsilon)\right]}e^{-iR(\varepsilon)},\quad
[K_m,\,D_0+\delta D(\varepsilon)]=0.
\label{eq:S2_normalform}
\end{equation}
The neutral generator is the group average of the diagonal layer over the on-site $m$-cycle,
\begin{equation}
D_0=\frac{1}{m}\sum_{j=0}^{m-1}K_m^{-j}H_zK_m^{j},\quad
\delta D(\varepsilon)=\frac{\varepsilon}{2}\,[D_0,G_{1,0}]+\mathcal{O}(\varepsilon^2).
\label{eq:S2_D0}
\end{equation}
We grade operators by the $K_m$-induced sectors as in Eqs.~\!\eqref{eq:S1_proj}-\eqref{eq:S1_chargecomm}: $X=\sum_q X_q$ with $K_m X_q K_m^\dagger=e^{2\pi iq/m}X_q$.

For a single on-site cycle with multiplicative imperfection,
\begin{equation}
K_m(\varepsilon)=K_m E(\varepsilon),\quad
E(\varepsilon)=\exp\!\big[-i\varepsilon G_1+\mathcal{O}(\varepsilon^2)\big],
\end{equation}
where $G_1$ is the linear error generator (for an angle-error like imperfection, $G_1=\theta_0 G$ on the active subspace). Decomposing $G_1=\sum_q G_{1,q}$ and expanding to first order gives
\begin{equation}
R(\varepsilon)=\varepsilon\sum_{q\neq 0}\left(G_{1,q}+\frac{i}{2}[D_0,G_{1,q}]\right)+\mathcal{O}(\varepsilon^2),
\label{eq:S2_R}
\end{equation}
so only the charged components ($q\neq 0$) contribute at leading order. By construction, $D_0+\delta D$ commute with $K_m$, and hence preserve the $\mathbb{Z}_m$ time charge, setting sector-preserving phases and slow envelopes without shifting the locked subharmonic frequencies. In contrast, $R(\varepsilon)$ carries nonzero time charge, mixes sectors, and can redistribute spectral weight among the harmonics and shift their centers.

\subsection{B3. Identical multiplicative imperfections}
\label{sec:B3_identical}

All numerics in the main text are performed with the same simple imperfection model: an identical multiplicative angle error applied to every instance of the on-site primitive that defines the carrier $K_m$. Concretely, if the ideal on-site $m$-cycle on site $i$ is generated by a Hermitian operator $G_i$ via
\begin{equation}
  K_{m,i} = e^{-i\theta_0 G_i},
\end{equation}
we implement an identical misrotation $\theta_0 \longrightarrow (1{+}\varepsilon)\,\theta_0$ for all sites and all steps. At the Floquet-step level this gives
\begin{equation}
  K_m(\varepsilon) = K_m E(\varepsilon), \quad
  E(\varepsilon) = \exp[-i\,\varepsilon G_1 + \mathcal{O}(\varepsilon^2)],\quad
  G_1 = \sum_i \theta_0 G_i,
\end{equation}
so that the static layer $e^{-iH_z}$ is unchanged and only the carrier acquires an identical angle error.

Because each $K_{m,i}$ is a function of $G_i$ alone, one has $[K_{m,i},G_i]=0$ on every site, and hence
\begin{equation}
  [K_m,G_1]=0, \quad K_m G_1 K_m^\dagger = G_1.
\end{equation}
Using the charge decomposition,
\begin{equation}
  G_1 = \sum_{q=0}^{m-1} G_{1,q}, \quad
  K_m G_{1,q} K_m^\dagger = e^{2\pi i q/m} G_{1,q},
\end{equation}
the commutation $[K_m,G_1]=0$ implies that $G_1$ is purely neutral:
\begin{equation}
  G_{1,0} = G_1, \quad G_{1,q} = 0 \quad \text{for all } q\neq 0.
\end{equation}
Inserting this into the linear-order expression for the charged remainder $R(\varepsilon)$ derived in Sec.~\!B2 (Eq.~\!\eqref{eq:S2_R}), we find that the entire $\mathcal{O}(\varepsilon)$ contribution cancels and
\begin{equation}
  R(\varepsilon) = \mathcal{O}(\varepsilon^2),
\end{equation}
for all protocols studied here (global and embedded $2T/3T$ as well as the $d{=}4,5$ examples). In the identical imperfection setting used in our simulations, the sector-mixing remainder is therefore parametrically subleading, and the leading architecture dependence in the observables is encoded in the neutral part $D(\varepsilon)$.

This conclusion is specific to the identical on-site structure of the imperfections. If the angle errors are non-uniform across sites or steps, for example
\begin{equation}
  G_1 = \sum_i \theta_0^{(i)} G_i + \Delta G_1,
\end{equation}
with site-dependent coefficients $\theta_0^{(i)}$ or additional contributions $\Delta G_1$ that do not commute with $K_m$, then $G_1$ generically acquires nonzero charged components $G_{1,q\neq 0}$. Substituting these into the same linear-order formula of Sec.~\!B2 yields a nonvanishing $\mathcal{O}(\varepsilon)$ contribution, and hence $R(\varepsilon) = \mathcal{O}(\varepsilon)$ in the generic nonidentical or step-dependent case. The identical multiplicative model used in our numerics should thus be viewed as a symmetry-enhanced limit of the general normal-form framework developed in Secs.~\!B1-B2.

\section{C. Observables}
\label{sec:C}

We measure stroboscopic expectation values at integer periods $t=nT$. The chain-averaged magnetization is
\begin{equation}
M_z(n)=\frac{1}{N}\sum_{i=0}^{N-1}\langle S_i^z(nT)\rangle,
\label{eq:S5_Mz}
\end{equation}
and analogous definitions apply to any on-site or few-body observable $O$. To isolate oscillatory contents, we remove the mean
\begin{equation}
\overline{M}_z=\frac{1}{N_t}\sum_{n=0}^{N_t-1}M_z(n),\quad \delta M_z(n)=M_z(n)-\overline{M}_z,
\label{eq:S5_DC}
\end{equation}
where $N_t=300$ is the number of recorded periods in our simulations. Unless stated, we report spectra for $\delta M_z$ using the full record (rectangular window), and the discrete frequency resolution is $\Delta f=1/N_t$. We then use the length-$N_t$ discrete Fourier transform
\begin{equation}
\widehat{x}(k)=\sum_{n=0}^{N_t-1} x(n)e^{-i2\pi kn/N_t},\quad k=0,1,\dots,N_t-1,
\label{eq:S5_DFT}
\end{equation}
with normalized frequencies $f_k=k/N_t$. The periodogram is
\begin{equation}
S^x(k)=\left|\widehat{x}(k)\right|^2.
\label{eq:S5_periodogram}
\end{equation}
For subharmonic diagnostics, we set $x=\delta M_z$ and identify the target index $k_m$ closest to $f=1/m$. We define the fractional Fourier weight at the subharmonic $1/m$ as
\begin{equation}
C_m=\frac{S_{\delta M_z}(k_m)}{\sum_{k=0}^{N_t-1}S_{\delta M_z}(k)}.
\label{eq:S5_Cm}
\end{equation}

\section{D. Circuits}
\label{sec:D}
This section summarizes the circuit architectures used in the main text.
\begin{figure}[htbp]
\centering
\includegraphics[width=0.45\linewidth]{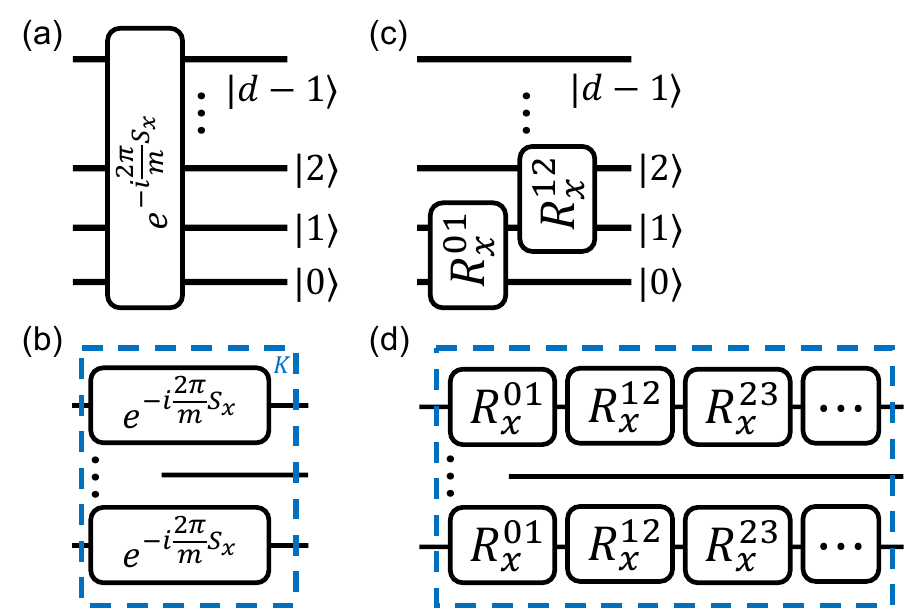}
\caption{General templates. (a,b) Single-site view: (a) all-level on-site $2\pi/m$ cycle acting on all $d$ levels. (b) Subspace-selective on-site cycle confined to an active subspace. Here, $R_x$ denotes $R_x(\pi)$. (c,d) Chain circuits view: (c) apply the single multilevel gate on each site. (d) Compile the on-site cycle with two-level primitives per site within the active subspace. The dashed blue box represents the kicking operation within one Floquet period.}
\label{fig:general_circ}
\end{figure}

\paragraph*{Fig.~\!\ref{fig:general_circ}.}
Panels (a,b) depict the single-site operator $K$ used in the Floquet step $U_F=K e^{-iH_z}$ at $\varepsilon=0$. In (a) the on-site gate is an all-level $2\pi/m$ cycle that acts on all $d$ local levels of the site. In (b) the on-site gate is subspace-selective: it cycles only an active subspace and acts as identity on the complementary levels. Since all angle for $R_x$ gates in this work equal to $\pi$, we henceforth abbreviate $R_x(\pi)$ as $R_x$ in figures. Panels (c,d) show the corresponding chain circuits built by applying the same on-site gate at each site: (c) uses the single multilevel gate per site, whereas (d) uses a short sequence of embedded two-level gates per site that realizes the same on-site cycle within the active subspace.

\begin{figure}[htbp]
\centering
\includegraphics[width=0.6\linewidth]{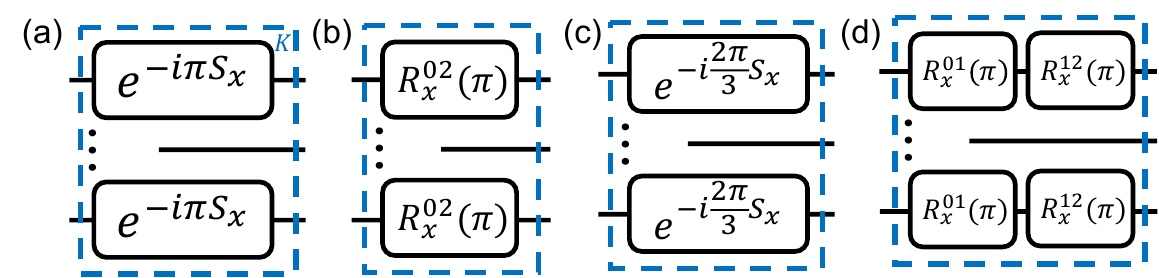}
\caption{$d{=}3$ (qutrit) circuits. (a) Global 2T: all-level on-site $\pi$ rotation on each site. (b) Embedded 2T: subspace-selective two-level $\pi$ rotation confined to $\{\ket{0},\ket{2}\}$ (level $\ket{1}$ inert). (c) Global 3T: all-level on-site $2\pi/3$ cycle over $\{\ket{0},\ket{1},\ket{2}\}$. (d) Embedded 3T: two sequential subspace-selective rotations on neighboring pairs that realize the three-cycle. The dashed blue box represents the kicking operation within one Floquet period.}
\label{fig:D3_circ}
\end{figure}

\paragraph*{Fig.\!~\ref{fig:D3_circ}.}
All panels show full-chain circuits for $d{=}3$ qudit kicks. (a) Global 2T: an all-level on-site $\pi$ rotation is applied on each site. (b) Embedded 2T: a two-level $\pi$ rotation confined to the active doublet $\{\ket{0},\ket{2}\}$ is applied per site, with $\ket{1}$ inert. (c) Global 3T: an all-level on-site $2\pi/3$ cycle over $\{\ket{0},\ket{1},\ket{2}\}$ is applied per site. (d) Embedded 3T: the three-cycle is realized by two sequential subspace-selective rotations on neighboring pairs (e.g., $\{0,1\}$ then $\{1,2\}$) per site.

\begin{figure}[htbp]
\centering
\includegraphics[width=0.65\linewidth]{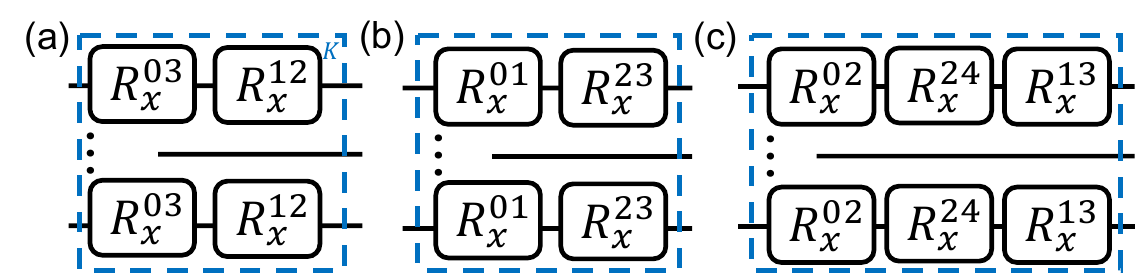}
\caption{$d{=}4$ and $d{=}5$ circuits. (a,b) Spin-$3/2$ (two doublets per site) with embedded 2T: (a) symmetry-paired partition $\{0,3\}\oplus\{1,2\}$. (b) Contiguous partition $\{0,1\}\oplus\{2,3\}$. In both, on-site two-level $\pi$ rotations act within the indicated doublets and leave the complementary levels inert. (c) Spin-2 mixed $3{:}2$: a trimer runs a three-cycle in parallel with a doublet flip, enabling concurrent $1/3$ and $1/2$ subharmonic channels within the same period. The dashed blue box represents the kicking operation within one Floquet period.}
\label{fig:D4_D5_circ}
\end{figure}

\paragraph*{Fig.~\!\ref{fig:D4_D5_circ}.}
All panels show subspace-selective (embedded) kicks. (a,b) Spin-$3/2$ ($d{=}4$): the local Hilbert space is partitioned into two doublets. In (a) the symmetry-paired partition $\{0,3\}\oplus\{1,2\}$ is used. In (b) the contiguous partition $\{0,1\}\oplus\{2,3\}$ is used. (c) Spin-2 ($d{=}5$) mixed trimer-doublet: a trimer $\{0,2,4\}$ is driven to realize a three-cycle in parallel with a doublet $\{1,3\}$ flip.

\section{E. Details of Spin-1 systems}
\label{sec:E}
\subsection{E1. Global vs.\ embedded in 2T: error structure and robustness}
\label{sec:E1}

We compare a global $\pi$ rotation on the full qutrit,
$K_{\mathrm{glob}}=e^{-i\pi(1+\varepsilon)S^x}$, with an embedded flip on a chosen doublet, e.g.\ $\{0,1\}$,
$K_{\mathrm{emb}}=e^{-i\pi(1+\varepsilon)\tau^{01}_x}\oplus I_{\{2\}}$.
The ideal $2T$ carriers are
\begin{equation}
\text{global: } K_2=e^{-i\pi S^x},\quad
\text{embedded: } K_2=e^{-i\pi \tau^{01}_x}\oplus I_{\{2\}},
\label{eq:S6_K2}
\end{equation}
and we take multiplicative imperfection, i.e.\ the angle error multiplies the same on-site generator that defines $K_2$, $\pi\mapsto\pi(1+\varepsilon)$. With the identical imperfection model of Sec.~\!B3 the leading error generator $G_1$ commutes with $K_2$ in both architectures and is therefore purely neutral. By the general analysis of Sec.~\!B, the charged remainder obeys $R(\varepsilon)=\mathcal{O}(\varepsilon^2)$ for both global and embedded 2T drives, their different robustness arises from the neutral sector $D(\varepsilon)$. 

From Eqs.~\!\eqref{eq:S2_normalform}-\eqref{eq:S2_D0}, we obtain,
\begin{equation}
D_0=\frac{1}{2}\sum_{j=0}^{1}K_2^{-j}H_zK_2^{j},\quad
\delta D(\varepsilon)=\frac{\varepsilon}{2}[D_0,G_1]+\mathcal{O}(\varepsilon^2),
\end{equation}
then:
\begin{enumerate}
  \item \emph{Global 2T ($K_2=e^{-i\pi S^x}$).} Since $K_2^{-1}S^zK_2=-S^z$, the on-site field terms in $H_z$ average to zero:
  \begin{equation}
  \frac{1}{2}\Big(S^z + K_2^{-1}S^z K_2\Big)=0,
  \end{equation}
  while the Ising couplings are invariant under the global flip. Hence
  \begin{equation}
  D_0^{\mathrm{glob}}=\tfrac{1}{2}\sum_{j=0}^{1}K_2^{-j}H_z K_2^{j}
  =\sum_i J^{z}_{i,i+1}S_i^z S_{i+1}^z,
  \quad
  \delta D^{\mathrm{glob}}(\varepsilon)=\tfrac{\varepsilon}{2}[D_0^{\mathrm{glob}},\pi S^x]+\mathcal{O}(\varepsilon^2).
  \label{eq:S6_D_global_re_new}
  \end{equation}
  Neutral dephasing is therefore generated by Ising couplings acting on the full three-level space via the commutator with $S^x$.

  \item \emph{Embedded 2T ($K_2=\tau_x^{01}\oplus I_{\{2\}}$).} Write $\Pi_{01}=\ket{0}\bra{0}+\ket{1}\bra{1}$, $\Pi_{2}=\ket{2}\bra{2}$, and $S^z=\mathrm{diag}(-1,0,1)$. A direct calculation gives
  \begin{equation}
  \frac{1}{2}\Big(S^z+K_2^{-1}S^z K_2\Big)=-\tfrac{1}{2}\Pi_{01}+1\cdot\Pi_{2},
  \label{eq:S6_avg_emb_Sz_only_re_new}
  \end{equation}
  so the on-site field term in $D_0$ becomes
  \begin{equation}
    \sum_i h_i S_i^z \longrightarrow
    \sum_i h_i\left(-\tfrac{1}{2}\Pi_{01,i}+1\cdot\Pi_{2,i}\right).
  \end{equation}
  Substituting into the definition of $D_0$ yields $D_0^{\mathrm{emb}}$ with the same Ising structure as in the global case plus these projectors, and
  \begin{equation}
    \delta D^{\mathrm{emb}}(\varepsilon)=\tfrac{\varepsilon}{2}[D_0^{\mathrm{emb}},\pi \tau_x^{01}]+\mathcal{O}(\varepsilon^2).
    \label{eq:S6_D_emb_re_new}
  \end{equation}
\end{enumerate}

The contrast between global and embedded 2T driving is now visible directly at the level of commutators. For the global drive, $D_0^{\mathrm{glob}}$ contains only $S^z_i S^z_{i+1}$ terms, and $[S_i^z S_{i+1}^z, S^x]$ acts nontrivially on all three local levels, so neutral dephasing of $M_z$ collects contributions from the entire qutrit. For the embedded drive, $D_0^{\mathrm{emb}}$ is block-diagonal and the active doublet is rotated by $\tau_x^{01}$ while the inert level is singled out by $\Pi_{2,i}$. Crucially,
\begin{equation}
  [\Pi_{2,i},\tau_x^{01}]=0,
\end{equation}
so all terms in $D_0^{\mathrm{emb}}$ that are proportional to $\Pi_{2,i}$ commute with the carrier and drop out of $[D_0^{\mathrm{emb}},\tau_x^{01}]$. Thus only the parts of the fields and couplings projected onto $\Pi_{01,i}$ contribute to the neutral commutator that governs the envelope of the active doublet, whereas contributions that live purely on the inert level $\ket{2}$ do not dephase the doublet coherence.

In summary, under the identical multiplicative imperfection model both global and embedded 2T protocols share the same leading charged remainder, $R(\varepsilon)=\mathcal{O}(\varepsilon^2)$, but differ in how $D(\varepsilon)$ acts on the observed subspace. The global drive generates neutral dephasing via interactions on the full three-level space, while the embedded drive restricts the effective neutral dephasing of the active doublet to the projection of disorder and couplings onto the $\{\ket{0},\ket{1}\}$ block. This difference underlies the slower decay of the subharmonic envelope observed for the embedded 2T implementation in the main text.

\subsection{E2. Why a global 3T step shows no observable line in $M_z$}
\label{sec:E2}

\paragraph*{Global 3T.}
Take $K_{\mathrm{glob}}=e^{-i\frac{2\pi}{3}(1+\varepsilon)S^x}$ with carrier $K_3=e^{-i\frac{2\pi}{3}S^x}$. For multiplicative imperfection $E(\varepsilon)=\exp[-i\varepsilon G_1+\mathcal{O}(\varepsilon^2)]$ one has $G_1=\tfrac{2\pi}{3}S^x$, which is neutral under the $\mathbb{Z}_3$ grading, so $G_{1,\pm1}=0$ and
\begin{equation}
V(\varepsilon)U_F(\varepsilon)V^\dagger(\varepsilon)=K_3\,e^{-i[D_0+\delta D(\varepsilon)]}\,e^{-iR(\varepsilon)},\quad
R(\varepsilon)=\mathcal{O}(\varepsilon^2).
\end{equation}
With $H_z=\sum_i J^{z}_{i,i+1}S_i^zS_{i+1}^z+\sum_i h_i S_i^z$, the neutral average is
\begin{equation}
D_0=\frac{1}{3}\sum_{j=0}^{2}K_3^{-j} H_z K_3^{j},\quad
\frac{1}{3}\sum_{j=0}^{2} K_3^{-j} S^z K_3^{j}=0,\quad
\frac{1}{3}\sum_{j=0}^{2} K_3^{-j} (S_i^z S_{i+1}^z) K_3^{j}
=\frac{1}{2}\!\left(S_i^z S_{i+1}^z+S_i^y S_{i+1}^y\right).
\end{equation}
Thus fields average out, while the couplings acquire an $S^yS^y$ term that does not commute with the charged parts of $M_z$. The evolution under $e^{-inD_0}$ therefore dephases the $1/3$ component already at $\varepsilon=0$, strongly suppressing a visible line in the chain-averaged $M_z$.

\paragraph*{Embedded 3T.}
Implement the three-cycle by two sequential embedded two-level $\pi$ rotations on neighboring pairs,
\begin{equation}
K_3^{\mathrm{emb}} \equiv R_x^{12}(\pi)\,R_x^{01}(\pi),
\end{equation}
which realizes $|0\rangle\!\to\!|1\rangle\!\to\!|2\rangle\!\to\!|0\rangle$. With the same multiplicative imperfection, the linear generator is neutral on the active trimer and $R(\varepsilon)=\mathcal{O}(\varepsilon^2)$ again. The neutral average
\begin{equation}
D_0^{\mathrm{emb}}=\frac{1}{3}\sum_{j=0}^{2}(K_3^{\mathrm{emb}})^{-j} H_z (K_3^{\mathrm{emb}})^{j}
\end{equation}
reduces to a level-diagonal average on the trimer: single-site fields drop out and the two-site part remains diagonal in the $\{\ket{0},\ket{1},\ket{2}\}$ basis. Equivalently, writing $\Pi_k=\ket{k}\!\bra{k}$, we have
\begin{equation}
\frac{1}{3}\sum_{j=0}^{2}(K_3^{\mathrm{emb}})^{-j}\big(S_i^zS_{i+1}^z\big)(K_3^{\mathrm{emb}})^{j}
= \sum_{k=0}^{2}\Pi_{k,i}\Pi_{k,i+1}-\frac{1}{3}I,
\end{equation}
so $D_0^{\mathrm{emb}}$ contains no off-diagonal terms. As a result, $M_z$ obtains only level-dependent phases from $D_0^{\mathrm{emb}}$, leading to milder neutral dephasing and a visible line near $1/3$ under the same conditions where the global 3T response is washed out.

\subsection{E3. Size effects: frequency offset and peak width}
\label{sec:E3}

\subsubsection{E3.1. Operational definitions.}
For each stroboscopic trace $M_z(n)$ at fixed imperfection $\varepsilon$, remove the mean $x(n)=M_z(n)-\overline{M_z}$. Let $N_t$ be the number of samples and define discrete frequencies $f_k=k/N_t$ (cycles per period). The power spectrum is
\begin{equation}
S_k=\Big|\sum_{n=0}^{N_t-1} x(n)\,e^{-i2\pi f_k n}\Big|^2.
\end{equation}
We analyze the spectral peak near $1/m$ in a narrow, fixed index window centered at the ideal index $k_0=\mathrm{round}(N_t/m)$. Choose a small half-bandwidth $h$ (a few bins, independent of $k_0$) and define
\begin{equation}
  \mathsf{W}=\{k|k_0-h\le k\le k_0+h\},\quad f_k=\frac{k}{N_t}.
\end{equation}
Let $f_{\mathrm{peak}}$ be the discrete frequency $f_k$ within $\mathsf{W}$ where $S_k$ attains its maximum. We report
\begin{equation}
\Delta f \equiv f_{\mathrm{peak}}-\frac{1}{m},\quad
\Gamma \equiv 2.355\,\sqrt{\sum_{k\in\mathsf{W}} p_k\,(f_k-f_{\mathrm{peak}})^2},\quad
p_k=\frac{S_k}{\sum_{j\in\mathsf{W}} S_j}.
\end{equation}
The factor $2.355$ converts the second central moment to a Gaussian full width at half maximum. Here $\Delta f$ is the offset of the peak center from $1/m$, and $\Gamma$ is a moment-based Gaussian FWHM proxy for the local peak width computed within the fixed window $\mathsf{W}$. In our data, $N_t$ is chosen to be the whole evolution time.

\subsubsection{E3.2. Toggle-frame expectations for size scaling.}
Within the dressed normal form of Eq.~\!\eqref{eq:S2_normalform}, the component at $1/m$ appears as a carrier fixed by $K_m$ multiplied by a slow envelope from $D(\varepsilon)\equiv D_0+\delta D(\varepsilon)$, while $R(\varepsilon)$ mixes time-charge sectors. Writing the dressed signal as a sum of local charged pieces,
\begin{equation}
x(n)\approx \Re\!\left[e^{i\frac{2\pi}{m} n}\,\frac{1}{N}\sum_{i=1}^N a_i(\varepsilon)\,e^{i\phi_i(\varepsilon)n}\right]+\text{(sector mixing from $R$)}.
\label{eq:E3_local_sum}
\end{equation}
Two consequences control the size dependence:

\smallskip
\noindent\emph{Neutral envelope and dephasing from $D(\varepsilon)$.}
The phases $\phi_i(\varepsilon)$ are set by on-site fields and nearby couplings after the $K_m$-average, hence they are determined by a finite neighborhood and their distribution does not change with $N$. The envelope in Eq.\!~\eqref{eq:E3_local_sum} is an empirical mean, $N^{-1}\sum_i e^{i\phi_i n}$, which converges for large $N$ to the disorder average $\langle e^{i\phi n}\rangle$ with fluctuations that decay as $N^{-1/2}$. Therefore, any finite width of the subharmonic peak that arises from neutral dephasing is set by the variance of these local phases, has an $N$-independent mean, and only exhibits $N^{-1/2}$ sample-to-sample variance.

\smallskip
\noindent\emph{Charged mixing from $R(\varepsilon)$.}
Sector mixing adds a further contribution to the shape of the subharmonic peak and sets any small center shift. Under the multiplicative imperfection used here, the leading charged term is suppressed so $R(\varepsilon)=\mathcal{O}(\varepsilon^2)$. Consequently
\begin{equation}
\Delta f(\varepsilon)=\mathcal{O}(\varepsilon^2)\quad\text{with an $N$-independent prefactor, and}\quad
\operatorname{std}[\Delta f]\sim N^{-1/2}.
\end{equation}
Summarizing, the width proxy $\Gamma(\varepsilon)$ is controlled by the variance of local neutral phases and is essentially insensitive to $N$, while $\Delta f(\varepsilon)$ is charge-controlled, small (quadratic in~$\varepsilon$), and its residual fluctuations shrink as $N^{-1/2}$.

\subsubsection{E3.3. Finite-size dependence}
\begin{figure}[htbp]
  \centering
  \includegraphics[width=0.8\linewidth]{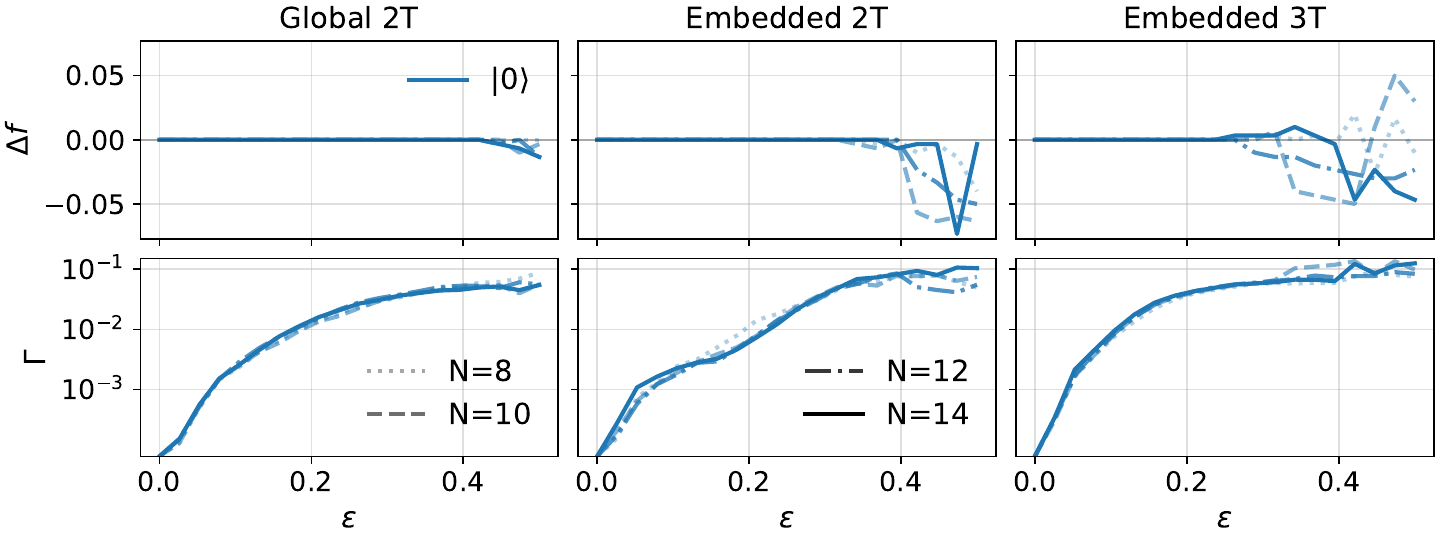}
  \caption{Finite-size trends using frequency offset $\Delta f$ and peak-width proxy $\Gamma$. For each trace we remove the mean, compute the discrete spectrum, take the most intense line within the fixed window $\mathsf{W}$ around $1/m$, set $\Delta f=f_{\mathrm{peak}}-1/m$, and estimate $\Gamma$ from the power-weighted second moment in $\mathsf{W}$ (Gaussian FWHM proxy). Top: $\Delta f$. Bottom: $\Gamma$. Columns: Global 2T, Embedded 2T, Embedded 3T. Centers remain locked over the shown range and $\Gamma$ grows smoothly with imperfection, with only weak $N$-dependence, in line with the locality and self-averaging analysis.}
  \label{fig:size_effect}
\end{figure}

We benchmark $d{=}3$ chains with system sizes $N=8,10,12,14$ across all three drive architectures (Global 2T, Embedded 2T, and Embedded 3T). As illustrated in Fig.~\!\ref{fig:size_effect}, the curves for both the frequency shift $\Delta f(\varepsilon)$ and the peak-width proxy $\Gamma(\varepsilon)$ effectively collapse onto similar trajectories, indicating negligible finite-size effects. This universality arises because the dominant mechanisms governing the spectral features, specifically the on-site neutral averaging and the nearest-neighbor charged couplings, are strictly local. Consequently, the characteristic dephasing scales are intensive, and increasing the chain length $N$ merely suppresses statistical fluctuations without altering the leading-order dynamics.

\subsection{E4. Phase diagrams and stability analysis}\label{subsec:cm_phase_diagrams}

Following the initial characterization of the subharmonic response for representative parameters in the main text, we further construct phase diagrams in multiple 2D parameter planes to quantitatively map the stability boundaries of DTC phases and provide general physical guidance. In terms of quantitative metrics and numerical settings, the subharmonic weight $C_m$ is employed as the order parameter to evaluate the strength of frequency locking. The computational model consists of an open-boundary spin chain with $N=10$ sites and local dimension $d=3$. The system evolves from a simple product state $\ket{\psi_0}=\ket{00\cdots 0}$ for $T=300$ full Floquet periods, with results averaged over $n_{\mathrm{seeds}}=30$ independent disordered realizations.

According to the Hamiltonian, the core control parameters governing the system dynamics include the interaction strength $J$, coupling disorder $W_J$, on-site disorder $W_h$, and kick imperfection $\varepsilon$. The stable distribution of DTC phases is then systematically examined across three typical parameter planes. Specifically, the first set of calculations fixes the disorder strengths at $(W_h,W_J)=(6,1)$ and sweeps the $(J,\varepsilon)$ space to investigate how interaction strength modulate non-equilibrium stability. Subsequently, the dynamical phase boundaries are explored in the $(W_h,\varepsilon)$ plane under the condition $(J,W_J)=(1,1)$ to quantify the effectiveness of on-site disorder in suppressing many-body heating and sector mixing. Finally, by selecting a typical non-zero perturbation $\varepsilon=0.2$ and keeping $W_J=1$ constant, the subharmonic response profile in the $(W_h,J)$ plane is mapped. This result provides a view of how the competition between disorder and interaction affects frequency-locking coherence at a specific imperfection level.

\begin{figure}[htbp]
    \centering
    \includegraphics[width=\textwidth]{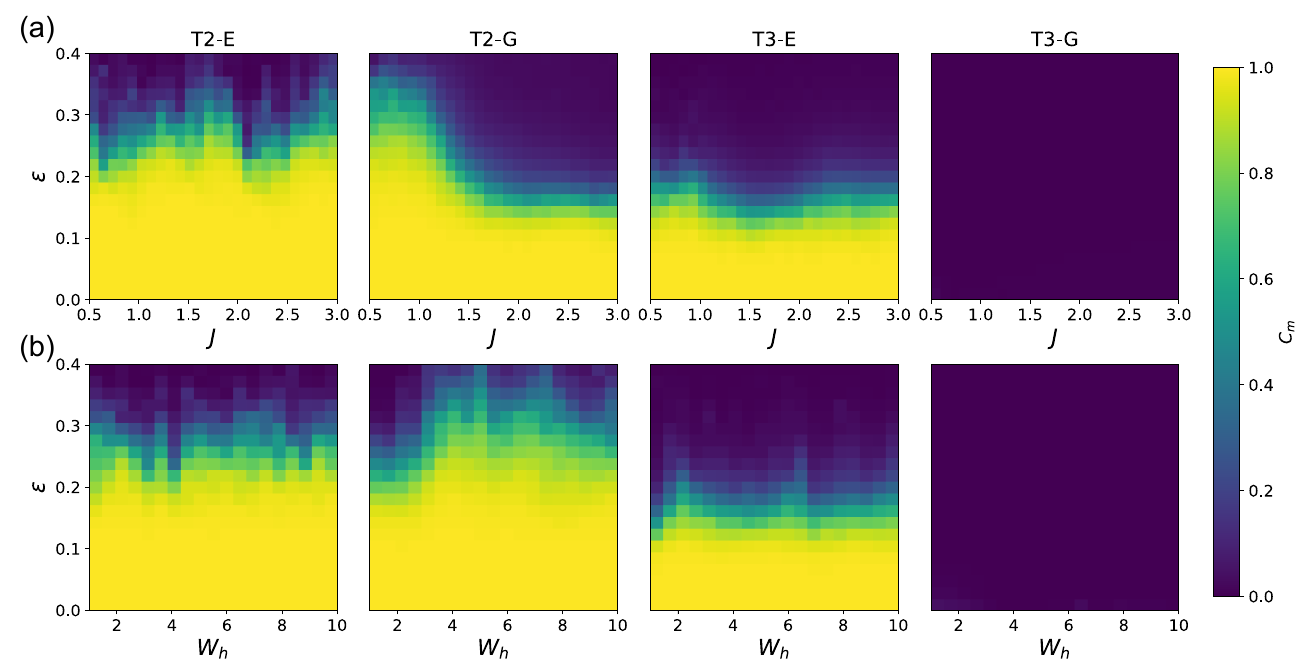}
    \caption{Discrete time crystal phase diagrams for spin-1 systems across different parameter planes ($N=10$). Calculations are performed under open boundary conditions for $T=300$ periods, with results averaged over $n_{\mathrm{seeds}}=30$ independent disordered realizations. The color scale represents the subharmonic weight $C_m$, where values closer to $1$ indicate a higher proportion of the target subharmonic component and stronger locking. (a) A $(J,\varepsilon)$ scan at fixed $(W_h,W_J)=(6,1)$ reveals that the embedded 2T kick remains robust against variations in J, while the global kick undergoes significant destabilization with increasing interaction strength. (b) A $(W_h,\varepsilon)$ scan at fixed $(J,W_J)=(1,1)$ demonstrates that the stability of the global kick relies on strong disorder $W_h$ to suppress many-body heating. Both (a) and (b) show that the global $3T$ kick fails to establish a dominant subharmonic response across the parameters.}
    \label{fig:cm_scheme12}
\end{figure}

Fig.~\!\ref{fig:cm_scheme12} highlights the distinct evolutionary trends in stability between different driving architectures. For the $2T$ DTC, the embedded kick exhibits a vast region of high $C_2$ in the $(J,\varepsilon)$ plane. The stability boundary shows minimal dependence on the interaction strength $J$, indicating that the system response remains firmly locked to the $1/2$ subharmonic for a reasonable range of $\varepsilon$. This phenomenon reflects the capability of the embedded structure to suppress perturbations, thereby delaying the breakdown of subharmonic locking across a wide parameter space. In contrast, the global kick architecture displays high sensitivity to $J$ in the $(J,\varepsilon)$ plane and requires larger on-site disorder $W_h$ to maintain stability in the $(W_h,\varepsilon)$ plane. This confirms that the global locking mechanism relies on the disorder field to suppress many-body mixing and heating channels. These results distinguish two stability mechanisms: the dynamical robustness of the driving structure and disorder-induced heating suppression. This interpretation is supported by the $3T$ DTC phase diagrams, while the embedded kick sustains visible 1/3 locking but shows increased sensitivity compared to the 2T case, as the imperfection $\varepsilon$ acts twice. Conversely, DTC signatures are entirely absent for the global 3T kick throughout the parameter space.

\begin{figure}[htbp]
    \centering
    \includegraphics[width=\textwidth]{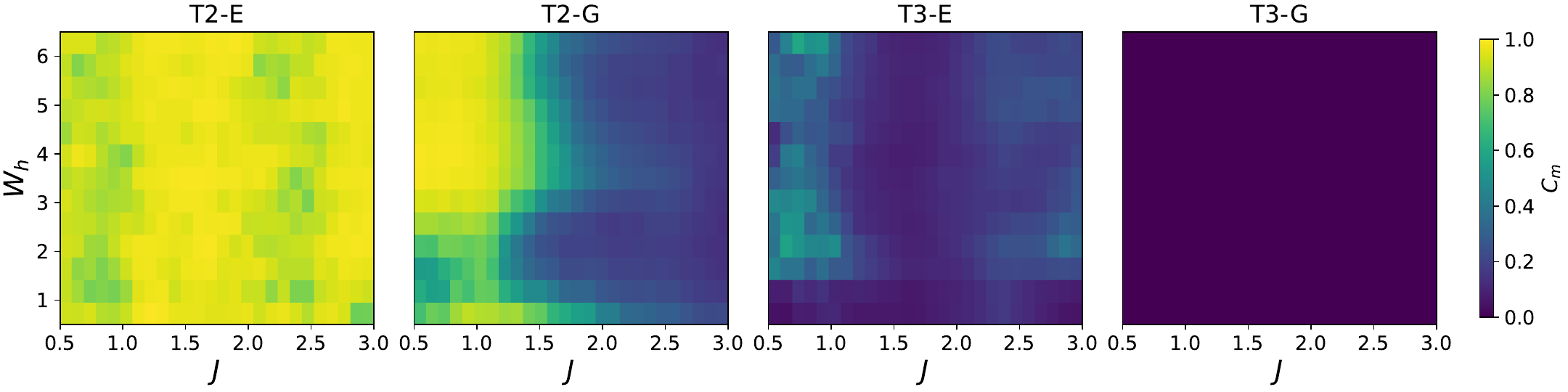}
    \caption{Discrete time crystal phase diagram in the $(W_h,J)$ plane at a fixed kick imperfection $\varepsilon=0.2$. Numerical settings other than the scanned parameters are identical to those in Fig.~\!\ref{fig:cm_scheme12} with $W_J=1$. The diagram illustrates the parameter dependence of different driving architectures at a specific imperfection level. The embedded $2T$ kick maintains high subharmonic weight over a broad parameter space, showing weak sensitivity to both disorder and interaction strength. In contrast, the locking capability of the global $2T$ kick degrades significantly under strong interactions and requires increased on-site disorder $W_h$ to suppress many-body heating. The global kick for 3T DTC again fails to exhibit any effective subharmonic dominance across the entire scanned plane.}
    \label{fig:cm_scheme3}
\end{figure}

To complement the stability boundary information at varying $\varepsilon$, Fig.~\!\ref{fig:cm_scheme3} presents the $C_m$ phase diagram in the $(W_h,J)$ plane with a fixed $\varepsilon=0.2$. The embedded $2T$ protocol maintains high $C_2$ values over a broad range of $(W_h,J)$, demonstrating relatively weak sensitivity to both parameters at this imperfection level. Conversely, the global $2T$ case shows a more parameter-dependent structure: increasing interaction strength typically reduces $C_2$, while increasing on-site disorder $W_h$ tends to enhancing the subharmonic locking by inhibiting heating and many-body mixing. The global $3T$ kick remains devoid of any significant $1/3$ coherent locking across the entire scanning plane.

In summary, these comprehensive phase diagrams provide a delineation of the stability boundaries for DTC phases within the $d=3$ qudit framework. By mapping the subharmonic response across the $(J, \varepsilon)$, $(W_h, \varepsilon)$, and $(W_h, J)$ parameter planes, we have achieved a complete characterization of this non-equilibrium phase. The results clearly distinguish two disparate stabilization regimes: while the global architecture relies heavily on strong disorder to mitigate many-body heating, the embedded protocol leverages its dynamical structure to suppress perturbations and maintain coherence. This systematic analysis confirms the reliability of our theoretical predictions and offers a clear operational window for observing stable qudit-based discrete time crystals in experimental platforms.

\section{F. Spin-3/2 Systems}
\label{sec:F}
\subsection{F1. Block averaging in the neutral generator}
\label{sec:F1}

Consider the embedded $2T$ protocol that partitions the on-site space into two disjoint doublets $A$ and $B$. Denote by $\Pi_{A,B}$ the projectors onto the two blocks and write the single-site operator $S^z$ restricted to block $b\in\{A,B\}$ as
\begin{equation}
S^z\big|_{b}=\mu_b\Pi_b+\delta_b \tau_b^z,
\quad
\mu_b=\tfrac{1}{2}(m_{b,1}+m_{b,2}),
\quad
\delta_b=\tfrac{1}{2}(m_{b,2}-m_{b,1}),
\label{eq:S7_block_decomp}
\end{equation}
where $m_{b,1}<m_{b,2}$ are the two $S^z$ eigenvalues in block $b$, and $\tau_b^{x,y,z}$ are Pauli operators acting within that doublet. The $2T$ carrier is $K_2=\tau_A^x\oplus\tau_B^x$, so $\tau_b^x\tau_b^z\tau_b^x=-\tau_b^z$ and $K_2 \Pi_b K_2^\dagger=\Pi_b$. Therefore, the group average that defines the neutral generator
\begin{equation}
D_0=\tfrac{1}{2}\big(H_z+K_2^{-1}H_z K_2\big)
\label{eq:S7_D0_def}
\end{equation}
acts on fields and couplings as follows.

\paragraph*{Single-site fields.}
Using Eq.~\!\eqref{eq:S7_block_decomp},
\begin{equation}
\tfrac{1}{2}\big(S^z+K_2^{-1}S^z K_2\big)
=\tfrac{1}{2}\sum_{b=A,B}\big(\mu_b\Pi_b+\delta_b\tau_b^z+\mu_b\Pi_b-\delta_b\tau_b^z\big)
=\mu_A\Pi_A+\mu_B\Pi_B.
\label{eq:S7_field_avg}
\end{equation}
Hence, the block means $\mu_b$ survive, while the contrast terms $\delta_b\tau_b^z$ cancel. For the symmetric pairing $\{0,3\}\oplus\{1,2\}$, one has $(m_A,m_B)=\{(-\tfrac{3}{2},+\tfrac{3}{2}),(-\tfrac{1}{2},+\tfrac{1}{2})\}\Rightarrow (\mu_A,\mu_B)=(0,0)$, so all on-site fields average out of $D_0$. For the contiguous pairing $\{0,1\}\oplus\{2,3\}$, one finds $(m_A,m_B)=\{(-\tfrac{3}{2},-\tfrac{1}{2}),(+\tfrac{1}{2},+\tfrac{3}{2})\}\Rightarrow (\mu_A,\mu_B)=(-1,+1)$. The quantity that controls neutral dephasing is the \emph{difference}
\begin{equation}
\Delta\mu\equiv \mu_B-\mu_A.
\end{equation}
If $\Delta\mu=0$ (even with $\mu_A=\mu_B\neq 0$), the field contribution reduces to a site-uniform multiple of the identity and produces only a global phase. When $\Delta\mu\neq 0$, the term $\mu_A\Pi_A+\mu_B\Pi_B$ acts as a block-dependent neutral field that dephases the subharmonic envelope in the presence of disorder or whenever the dynamics sample both blocks.

\paragraph*{Nearest-neighbor couplings.}
On two sites $(i,i{+}1)$ with block labels $(b,b')$,
\begin{equation}
\begin{aligned}
S_i^z S_{i+1}^z&=\big(\mu_b\Pi_{b,i}+\delta_b\tau_{b,i}^z\big)\big(\mu_{b'}\Pi_{b',i+1}+\delta_{b'}\tau_{b',i+1}^z\big)\\
&=\mu_b\mu_{b'}\Pi_{b,i}\Pi_{b',i+1}+\mu_b\delta_{b'}\Pi_{b,i}\tau_{b',i+1}^z+\delta_b\mu_{b'}\tau_{b,i}^z\Pi_{b',i+1}+\delta_b\delta_{b'} \tau_{b,i}^z\tau_{b',i+1}^z.
\end{aligned}
\end{equation}
Under conjugation by $K_2$ both $\tau^z$ factors flip sign, so the linear terms cancel on averaging while the quadratic term is invariant:
\begin{equation}
\tfrac{1}{2}\big(S_i^z S_{i+1}^z+K_2^{-1}S_i^z S_{i+1}^z K_2\big)
=\mu_b\mu_{b'}\Pi_{b,i}\Pi_{b',i+1}
+\delta_b\delta_{b'} \tau_{b,i}^z\tau_{b',i+1}^z.
\label{eq:S7_coupling_avg}
\end{equation}
Thus $D_0$ contains only a block-constant offset from $\mu_b\mu_{b'}\Pi_{b,i}\Pi_{b',i+1}$ and a neutral Ising term $\tau^z\tau^z$ within and across blocks, while all block-odd pieces are removed. For the symmetric pairing ($\mu_A=\mu_B=0$, so $\Delta\mu=0$), the single-site neutral field vanishes and neutral dynamics within each doublet are generated solely by the $\tau^z\tau^z$ term, which preserves doublet populations and leads to relatively weak neutral dephasing. In contrast, the contiguous pairing has $\Delta\mu\neq 0$, producing block-dependent neutral phases that more strongly dephase the subharmonic envelope.

\subsection{F2. Composite depth and the scaling of the charged residual}
\label{sec:F2}
Write the dressed period as
\begin{equation}
V(\varepsilon)U_F(\varepsilon)V^\dagger(\varepsilon)
=K_2\,e^{-i[D_0+\delta D(\varepsilon)]}\,e^{-iR(\varepsilon)},\quad
\|R(\varepsilon)\|=\mathcal{O}(\varepsilon^\alpha).
\label{eq:S7_normalform}
\end{equation}
For the embedded $2T$ protocol on $d=4$, the ideal on-site kick is a direct sum of disjoint $\pi$ rotations on the two doublets,
\begin{equation}
K_m=\exp[-i\pi \tau_A^x]\oplus\exp[-i\pi \tau_B^x],
\label{eq:S7_Kid}
\end{equation}
so, with multiplicative imperfection, $E(\varepsilon)=\exp[-i\varepsilon G_1+\mathcal{O}(\varepsilon^2)]$ where $G_1=\pi(\tau_A^x\oplus\tau_B^x)$ is neutral under $K_2$ ($G_{1,q\neq 0}=0$). Hence, the linear charged sector is absent and minimal implementations already achieve
\begin{equation}
R(\varepsilon)=\mathcal{O}(\varepsilon^2).
\label{eq:S7_alpha_min}
\end{equation}
At $\mathcal{O}(\varepsilon^2)$, charged contributions originate from nested commutators generated by the BCH/Magnus expansion of the compiled steps together with $H_z$ (e.g., terms of the schematic form $[G_1,[H_z,G_1]]$ and $[G_a,[H_z,G_b]]$ projected onto the sectors with $q\neq 0$). 

\subsection{F3. Higher-order subharmonic responses: cyclic $4T$ and trimer $3T$}
\label{sec:F3}
\begin{figure}[htbp]
  \includegraphics[width=0.4\linewidth]{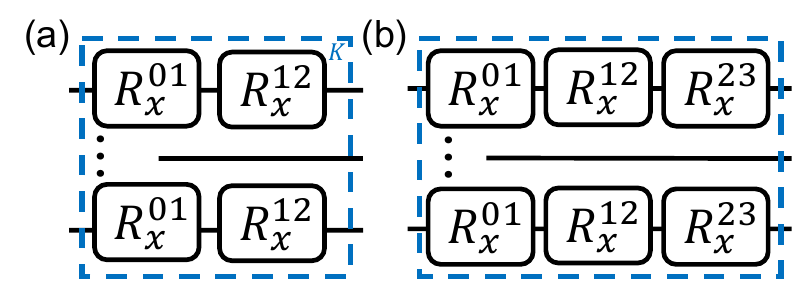}
  \caption{Circuits for $d=4$ systems. (a) Trimer 3T. (b) Cyclic $4T$. The dashed blue box represents the kicking operation within one Floquet period.}
  \label{fig:d4_T3_T4_circ}
\end{figure}

\begin{figure}[htbp]
\centering
\includegraphics[width=0.6\linewidth]{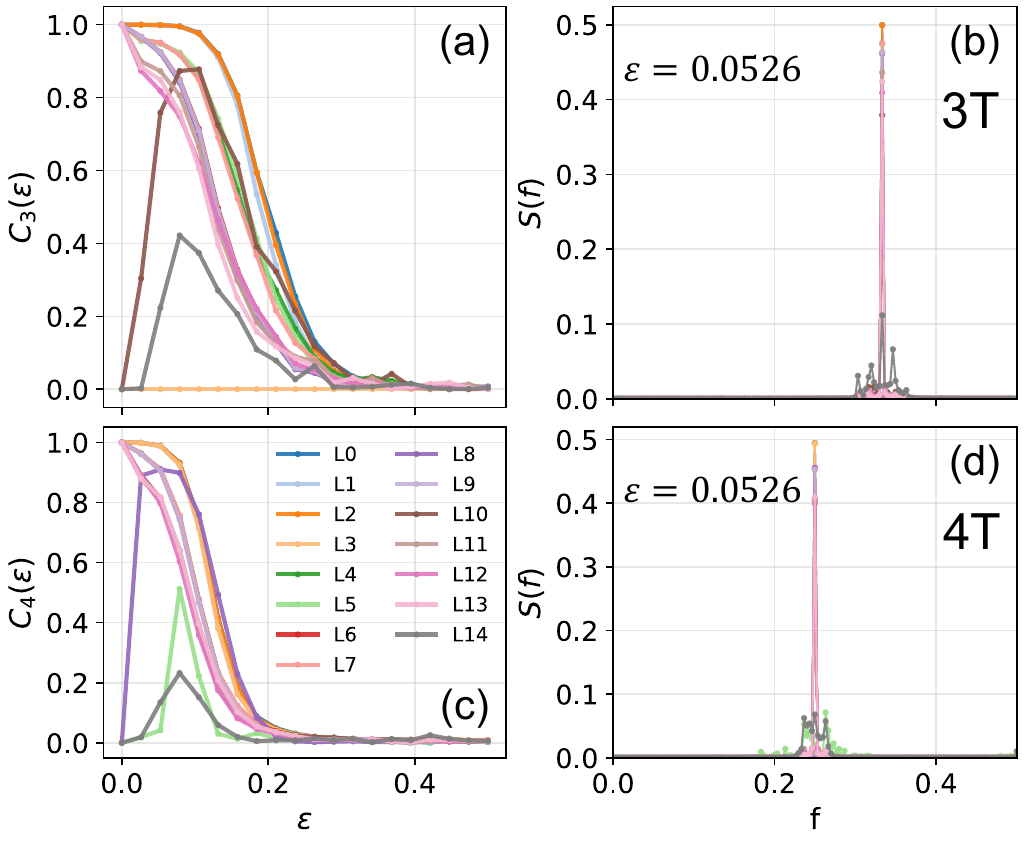}
\caption{Spin-$3/2$ responses for fifteen preparations L0-L14 (state map in Table~\!\ref{tab:Slabels}) with $N=6$. 
(a) $C_3(\varepsilon)$ for the trimer-$3T$ drive: most states lock robustly at $1/3$. L3 ($\ket{3}$) is inert by construction, while L10 and L14 show reduced contrast due to larger neutral support. 
(b) Representative spectra at $\varepsilon=0.0526$ displaying a sharp $1/3$ peak. 
(c) $C_4(\varepsilon)$ for the cyclic-$4T$ drive evaluated with the charged probe $\Omega_4=\mathrm{diag}(1,i,-1,-i)$, which isolates the $q{=}1$ sector and yields a clean $1/4$ line. Preparations L5 ($\ket{0}{+}\ket{2}$), L8 ($\ket{1}{+}\ket{3}$), and L14 are less favorable, consistent with their time-charge projections.}
\label{fig:d4_34}
\end{figure}

\paragraph*{Trimer $3T$ (active block $\{0,1,2\}$).} 
We synthesize a one-step 3-cycle with two adjacent two-level rotations (the compilation circuit is shown in Fig.~\!\ref{fig:d4_T3_T4_circ}(a)) on the active trimer, leaving $\ket{3}$ inert. Dressing yields
\begin{equation}
V(\varepsilon)U_F(\varepsilon)V^\dagger(\varepsilon)
=K_3e^{-i[D_0+\delta D(\varepsilon)]}e^{-iR(\varepsilon)},\quad K_3^3=I.
\label{eq:E3_nf3}
\end{equation}
Projecting $M_z$ onto the active block gives $\big(M_z^{\{0,1,2\}}\big)_{q=1}\neq 0$, so the chain-averaged signal contains a locked $1/3$ component. Population on $\ket{3}$ contributes only to the neutral sector and reduces contrast but does not shift the carrier. This explains Fig.~\!\ref{fig:d4_34}(a, b): basis states in the trimer and most superpositions yield a sharp $1/3$ peak, L3 is inert, and nearly uniform states over the trimer (e.g., L10) or over all four levels (L14) dilute the charged projection and lower $C_3$.

\paragraph*{Cyclic $4T$ (all four levels).}
A 4-cycle is compiled from three adjacent two-level rotations within the on-site quadruplet (the circuit is shown in Fig.~\!\ref{fig:d4_T3_T4_circ}(b)), with dressed form
\begin{equation}
V(\varepsilon)U_F(\varepsilon)V^\dagger(\varepsilon)
=K_4 e^{-i[D_0+\delta D(\varepsilon)]}e^{-iR(\varepsilon)},\quad K_4^4=I.
\label{eq:E3_nf4}
\end{equation}
Under the $\mathbb{Z}_4$ grading, $M_z=\sum_{q=0}^{3}(M_z)_q$ generally has $(M_z)_{q=1,2,3}\neq 0$. In particular, the equally spaced $S^z$ eigenvalues $(-\tfrac{3}{2},-\tfrac{1}{2},\tfrac{1}{2},\tfrac{3}{2})$ imply an anti-periodicity $x_{n+2}=-x_n$ for the $q=2$ sector, so a chain-averaged $M_z$ spectrum naturally mixes a $1/2$ harmonic with the $1/4$ line. To isolate the $q=1$ content, we use the charged probe
\begin{equation}
\Omega_4=\mathrm{diag}(1,i,-1,-i),\quad K_4\Omega_4K_4^\dagger=i\Omega_4,
\label{eq:E3_Omega4}
\end{equation}
for which $(\Omega_4)_{q=1}=\Omega_4$ and $(\Omega_4)_{q\neq 1}=0$. Fig.~\!\ref{fig:d4_34}(c, d) shows robust $1/4$ locking under $\Omega_4$ for most preparations. Opposite-pair superpositions L5 and L8 project dominantly into the $q=2$ sector of $M_z$ and have weak overlap with $q=1$, while the fully symmetric L14 has vanishing charged projection. Using $\Omega_4$ isolates the  $1/4$ channel, while spectra with $M_z$ display the expected mixture of $1/4$ and $1/2$ components consistent with the sector decomposition above.

\begin{table}[htbp]
\centering
\begin{tabular}{lll}
L0: $\ket{0}^{\otimes N}$ &
L1: $\ket{1}^{\otimes N}$ &
L2: $\ket{2}^{\otimes N}$ \vspace{0.2cm}\\\vspace{0.2cm}
L3: $\ket{3}^{\otimes N}$ &
L4: $\left(\frac{\ket{0}+\ket{1}}{\sqrt{2}}\right)^{\otimes N}$ &
L5: $\left(\frac{\ket{0}+\ket{2}}{\sqrt{2}}\right)^{\otimes N}$\\\vspace{0.2cm}
L6: $\left(\frac{\ket{1}+\ket{2}}{\sqrt{2}}\right)^{\otimes N}$ &
L7: $\left(\frac{\ket{2}+\ket{3}}{\sqrt{2}}\right)^{\otimes N}$ &
L8: $\left(\frac{\ket{1}+\ket{3}}{\sqrt{2}}\right)^{\otimes N}$ \\\vspace{0.2cm}
L9: $\left(\frac{\ket{3}+\ket{0}}{\sqrt{2}}\right)^{\otimes N}$ &
L10: $\left(\frac{\ket{0}+\ket{1}+\ket{2}}{\sqrt{3}}\right)^{\otimes N}$ &
L11: $\left(\frac{\ket{1}+\ket{2}+\ket{3}}{\sqrt{3}}\right)^{\otimes N}$ \\\vspace{0.2cm}
L12: $\left(\frac{\ket{2}+\ket{3}+\ket{0}}{\sqrt{3}}\right)^{\otimes N}$ &
L13: $\left(\frac{\ket{3}+\ket{0}+\ket{1}}{\sqrt{3}}\right)^{\otimes N}$ &
L14: $\left(\frac{\ket{0}+\ket{1}+\ket{2}+\ket{3}}{\sqrt{4}}\right)^{\otimes N}$ \\
\end{tabular}
\caption{Label map and single-site states for the $d=4$ initial state preparations. }
\label{tab:Slabels}
\end{table}

\subsection{F4. Two-qubit baselines for the doublet partition scheme}

\subsubsection{F4.1 Motivation, framework, and exact mappings}
\label{ssec:F4_framework}
We aim to separate the roles of local Hilbert space dimension and ZZ coupling geometry in frequency locking and the stability of the subharmonic response. For the spin-3/2 chain we consider the contiguous doublet partition $\{0,1\}\oplus\{2,3\}$, where the embedded 2T on-site kick factorizes into two commuting $\pi$ rotations. This enables an exact site-wise mapping to two qubits and a controlled baseline that alters only the ZZ connectivity while keeping the drive and on-site fields fixed. By contrast, the symmetric split $\{0,3\}\oplus\{1,2\}$ does not reduce to a single-leg rotation under the same encoding and its neutral group average removes on-site fields, which would intertwine drive resources with geometry. The contiguous case thus offers the cleanest setting for the isolation task pursued in this subsection. The single-step kick is 
\begin{equation}
U_{\mathrm{kick}}^{(4)}(\varepsilon)
=
\exp\left[-\frac{i\pi}{2}(1{+}\varepsilon)X_{01}\right]
\exp\left[-\frac{i\pi}{2}(1{+}\varepsilon)X_{23}\right],
\quad
X_{jk}=\ketbra{j}{k}+\ketbra{k}{j}.
\end{equation}

\paragraph*{Exact two-qubit encoding (\textsc{enc}).}
Define $W:\mathbb{C}^4\to\mathbb{C}^2_A\otimes\mathbb{C}^2_B$ by
\begin{equation}
|0\rangle\mapsto|00\rangle,\quad |1\rangle\mapsto|01\rangle,\quad
|2\rangle\mapsto|10\rangle,\quad |3\rangle\mapsto|11\rangle,
\end{equation}
with $A$ the high bit and $B$ the low bit. Then
\begin{equation}
WS^{z}W^\dagger = -Z_A - \frac{1}{2} Z_B,
\end{equation}
\begin{equation}
W^{\otimes 2}\left(S^{z}\otimes S^{z}\right)\left(W^\dagger\right)^{\otimes 2}
= Z_A Z_A + \frac{1}{2}\left(Z_A Z_B + Z_B Z_A\right) + \frac{1}{4} Z_B Z_B,
\end{equation}
so the diagonal layer becomes
\begin{equation}
H_z^{(\textsc{enc})}
=
\sum_{i} h_i\left(-Z_{A,i}-\frac{1}{2} Z_{B,i}\right)
+\sum_i J_i\left[
Z_{A,i}Z_{A,i+1}
+\frac{1}{2}\left(Z_{A,i}Z_{B,i+1}+Z_{B,i}Z_{A,i+1}\right)
+\frac{1}{4} Z_{B,i}Z_{B,i+1}\right].
\end{equation}
Since $X_{01}$ and $X_{23}$ commute, their sum maps to a single $B$-leg rotation:
\begin{equation}
X_{01}+X_{23} \xrightarrow{W} X_B,
\quad
U_{\mathrm{kick}}^{(4)}(\varepsilon)\xrightarrow{W}
\exp\left[-\frac{i\pi}{2}(1{+}\varepsilon)X_B\right]
= \prod_{i} R_x^{(B_i)}\big(\pi(1{+}\varepsilon)\big).
\end{equation}
Hence for any $\{h_i,J_i\}$ and any $\varepsilon$,
\begin{equation}
\big(W^{\otimes N}\big)U_F^{(4)}(\varepsilon)\big(W^\dagger\big)^{\otimes N}
=
U_F^{(\textsc{enc})}(\varepsilon)
\equiv 
\left[\prod_{i} R_x^{(B_i)}(\pi(1{+}\varepsilon))\right]e^{-i H_z^{(\textsc{enc})}},
\end{equation}
and observables related by $W$ (for example $S^z\mapsto -Z_A-\tfrac12 Z_B$) exhibit identical stroboscopic signals in the qudit and \textsc{enc} descriptions.

\paragraph*{Plain two-qubit baseline (\textsc{plain}).}
To isolate the impact of $ZZ$ geometry, we keep the same on-site fields and the same $B$-only kick as in \textsc{enc}, but use leg-preserving couplings with a tunable cross-leg weight:
\begin{equation}
H_z^{(\textsc{plain})}
=
\sum_{i} h_i\left(-Z_{A,i}-\frac{1}{2} Z_{B,i}\right)
+\sum_i J_i\left[
Z_{A,i}Z_{A,i+1}+Z_{B,i}Z_{B,i+1}
+\lambda\left(Z_{A,i}Z_{B,i+1}+Z_{B,i}Z_{A,i+1}\right)\right],
\quad \lambda\in[0,1],
\end{equation}
interpolating from decoupled legs $(\lambda=0)$ to the encoded value $(\lambda=\tfrac12)$. The Floquet step is
\begin{equation}
U_F^{(\textsc{plain})}(\varepsilon)
= \left[\prod_{i} R_x^{(B_i)}(\pi(1{+}\varepsilon))\right] e^{-i H_z^{(\textsc{plain})}}.
\end{equation}
By construction, \textsc{plain} differs from the qudit and \textsc{enc} models only in the $ZZ$ connectivity. In the baseline discussion below we use the plain case with $\lambda=0$, and the encoded geometry corresponds to $\lambda=\tfrac12$. 

\subsubsection{F4.2 Two-period normal form and even/odd structure}

Let the ideal $2T$ carrier be
\begin{equation}
K_0=\prod_i R_x^{(B_i)}(\pi),
\end{equation}
which satisfies
\begin{equation}
K_0^\dagger Z_A K_0 = Z_A,\quad K_0^\dagger Z_B K_0 = - Z_B.
\end{equation}
For any operator $X$, define the decomposition into parts that are invariant (``even'') or change sign (``odd'') under conjugation by $K_0$:
\begin{equation}
X^{(+)}=\tfrac{1}{2}\big(X+K_0^\dagger X K_0\big),\quad
X^{(-)}=\tfrac{1}{2}\big(X-K_0^\dagger X K_0\big).
\end{equation}
The leading two-period average is then
\begin{equation}
D_0=\tfrac{1}{2}\!\left[H_z^{(+)}+K_0^\dagger H_z^{(+)}K_0\right]=H_z^{(+)}.
\end{equation}

\paragraph*{Even/odd content of $H_z$ for \textsc{enc} and \textsc{plain}.}
Under the site-wise encoding (Sec.~\!F4.1), $S^z \mapsto -Z_A-\tfrac12 Z_B$, so
\begin{equation}
S_i^z S_{i+1}^z \ \mapsto\ 
Z_{A,i}Z_{A,i+1}+\tfrac12\!\left(Z_{A,i}Z_{B,i+1}+Z_{B,i}Z_{A,i+1}\right)+\tfrac14\,Z_{B,i}Z_{B,i+1}.
\end{equation}
Hence for the encoded geometry (\textsc{enc})
\begin{equation}
H_z^{(\textsc{enc})}
=
\underbrace{\sum_i h_i\!\left(-Z_{A,i}-\tfrac12 Z_{B,i}\right)}_{\text{even}}
+
\underbrace{\sum_i J_i\!\left[Z_{A,i}Z_{A,i+1}+\tfrac14 Z_{B,i}Z_{B,i+1}\right]}_{\text{even}}
+
\underbrace{\sum_i J_i\,\tfrac12\!\left(Z_{A,i}Z_{B,i+1}+Z_{B,i}Z_{A,i+1}\right)}_{\text{odd}}.
\label{eq:F42_enc_evenodd}
\end{equation}
Thus the cross-leg couplings are odd and drop out of $D_0$, while $Z_AZ_A$ and $Z_BZ_B$ survive in $D_0$ with weights $1$ and $1/4$, respectively.
For the leg-preserving baseline (\textsc{plain}) at $\lambda=0$ one has $H_z^{(-)}\equiv 0$ and
\begin{equation}
D_0^{(\textsc{plain})}=H_z^{(\textsc{plain})}=\sum_i h_i\!\left(-Z_{A,i}-\tfrac12 Z_{B,i}\right)+\sum_i J_i\!\left(Z_{A,i}Z_{A,i+1}+Z_{B,i}Z_{B,i+1}\right),
\end{equation}
so both legs enter $D_0$ with unit weight and no odd part is present.

\paragraph*{Multiplicative imperfection and its consequences.}
With a small multiplicative imperfection aligned to the $B$-leg rotation,
\begin{equation}
K_m(\varepsilon)=\prod_i R_x^{(B_i)}\!\big(\pi(1{+}\varepsilon)\big)
=\exp\left[-\frac{i\pi\varepsilon}{2}\sum_i X_{B,i}\right]K_0,
\end{equation}
a two-period BCH/Magnus expansion yields
\begin{equation}
\bar H=D_0+\varepsilon D_1+\mathcal{O}(\varepsilon^2),\quad
D_1\sim \frac{\pi}{2\Omega}\,[H_z^{(-)},\sum_i X_{B,i}] \,+\,\text{commutators with } H_z^{(+)} ,
\end{equation}
where $\Omega$ denotes the drive angular frequency. In this setting the linear error generator commutes with $K_0$ on the active block, so the charged remainder in the normal form satisfies $R(\varepsilon)=\mathcal{O}(\varepsilon^2)$ and the subharmonic line center remains locked. The term $[H_z^{(-)},\sum_i X_{B,i}]$ identifies how odd content of $H_z$ feeds the leading second-order charged items in the two-period analysis: when $H_z^{(-)}\neq 0$ (as in \textsc{enc}), it enhances the coefficient governing this $\mathcal{O}(\varepsilon^2)$ sensitivity. When $H_z^{(-)}=0$ (as in \textsc{plain} at $\lambda=0$), this contribution is absent and the charged channel first appears through higher commutators. The commutators with $H_z^{(+)}$ contribute only to neutral corrections and therefore affect the envelope but not the locked center.

Under the same imperfection, both \textsc{enc} and \textsc{plain} exhibit $R(\varepsilon)=\mathcal{O}(\varepsilon^2)$ and center locking. 
The decisive difference is that \textsc{enc} contains an odd part $H_z^{(-)}$ arising from cross-leg couplings [Eq.~\!\eqref{eq:F42_enc_evenodd}], which increases the leading neutral and charged coefficients entering $\bar H$ and thus accelerates the decay of the subharmonic response. In \textsc{plain} at $\lambda=0$ one has $H_z^{(-)}=0$, so this contribution is absent and the response is more stable. The different $Z_B Z_B$ weights in $D_0$ (unit in \textsc{plain} versus $1/4$ in \textsc{enc}) merely rescale the neutral envelope and do not change the perturbative order of the charged items. Consequently, within this common normal-form framework and for multiplicative imperfection, the leg-preserving geometry (\textsc{plain}, $\lambda=0$) is more robust than the encoded geometry (\textsc{enc}, $\lambda=\tfrac12$): both keep the line center fixed, but \textsc{plain} exhibits a slower decay and a longer imperfection plateau. 

\subsubsection{F4.3 Numerical benchmarks and interpretation}
In the contiguous partition the qudit curves match \textsc{enc} point by point, confirming the site-wise encoding. The \textsc{plain} baseline removes cross-leg couplings and increases the effective $Z_BZ_B$ weight in $D_0$, and it is therefore more robust. Because the drive primitives and on-site fields are fixed while only the ZZ connectivity is changed, this ordering reflects coupling geometry rather than local dimension. Beyond this baseline, qudits offer capabilities not captured by two-qubit reductions, including subspace-selective compilation, parallel subharmonic channels within a single drive, confinement of the drive to an active block that suppresses charged mixing into spectator levels, direct control of the envelope through block averaging in $D_0$, and richer period structures enabled by native multilevel selection rules. These features provide the design flexibility used in this work. 

\begin{figure}[htbp]
\centering
\includegraphics[width=0.7\linewidth]{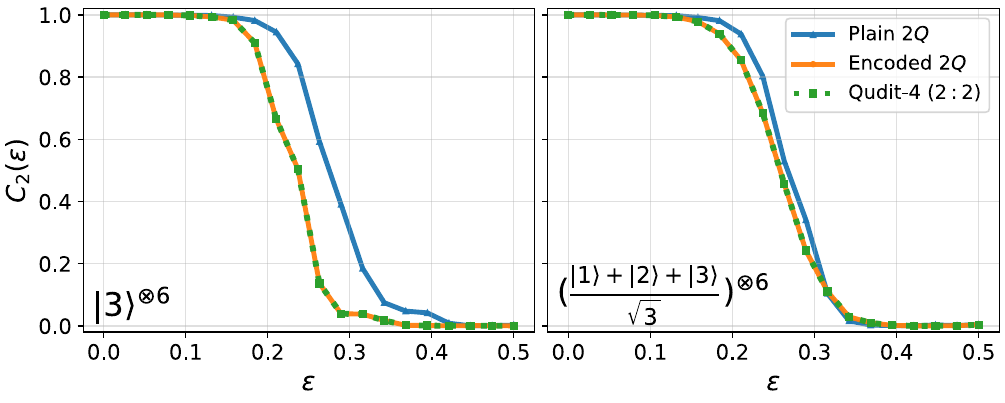}
\caption{Subharmonic weight $C_2(\varepsilon)$ for the $d=4$ chain under the contiguous doublet partition $\{0,1\}\oplus\{2,3\}$ and comparison to two-qubit baselines (\textsc{enc} and \textsc{plain} with total qubits $2N=20$). Initial states are $|3\rangle^{\otimes N}$ and $\big[(|1\rangle{+}|2\rangle{+}|3\rangle)/\sqrt{3}\big]^{\otimes N}$. The qudit data coincide with the exact encoding \textsc{enc}. The \textsc{plain} baseline differs only by the $ZZ$ geometry and shows a higher and longer plateau with a sharper drop, consistent with the even/odd analysis.}
\label{fig:D4_pattern}
\end{figure}

\section{G. Spin-2: trimer-doublet mixed protocol}
\label{sec:G}

\subsection{G1. Block structure, carrier, and imperfections}
\label{ssec:G1}

Partition each spin-2 site into a trimer $\mathcal{T}=\{0,2,4\}$ and a doublet $\mathcal{B}=\{1,3\}$ with single-site projectors
\begin{equation}
\Pi_{\mathcal{T}}=\ketbra{0}{0}+\ketbra{2}{2}+\ketbra{4}{4},\quad
\Pi_{\mathcal{B}}=\ketbra{1}{1}+\ketbra{3}{3},\quad
\Pi_{\mathcal{T}}+\Pi_{\mathcal{B}}=I.
\label{eq:S8_proj}
\end{equation}
The embedded kick compiles a one-step 3-cycle on $\mathcal{T}$ and a one-step 2-cycle on $\mathcal{B}$ within the same Floquet period, so the single-site carrier is block diagonal,
\begin{equation}
K_{\mathrm{site}}=K_3\oplus K_2,\quad K_3^3=K_2^2=I,
\label{eq:S8_Psite}
\end{equation}
and the many-body carrier is the tensor product $K=\bigotimes_i K^{(i)}$. Because the static layer $H_z$ in Eq.~\!\eqref{eq:S1_Hz} is diagonal in the computational basis, it preserves the on-site decomposition $\Pi_{\mathcal{T}}\oplus\Pi_{\mathcal{B}}$. In the dressed normal form one has
\begin{equation}
V(\varepsilon)U_F(\varepsilon)V^\dagger(\varepsilon)=
K e^{-i\left[D_0+\delta D(\varepsilon)\right]} e^{-iR(\varepsilon)},\quad [K,D_0+\delta D(\varepsilon)]=0,
\label{eq:S8_nf}
\end{equation}
where $D_0$ is the neutral average and $\delta D(\varepsilon)$ its first-order correction. We adopt the identical multiplicative imperfection model used throughout,
\begin{equation}
K(\varepsilon)=KE(\varepsilon),\quad
E(\varepsilon)=\exp\left[-i\varepsilon G_1+\mathcal{O}(\varepsilon^2)\right],
\label{eq:S8_err_model}
\end{equation}
in which the angle error multiplies the same on-site generators that define $K_3$ and $K_2$. For the embedded mixed trimer-doublet kick, $G_1$ is a sum of block-local generators on $\mathcal{T}$ and on $\mathcal{B}$ and is therefore block diagonal:
\begin{equation}
\Pi_{\mathcal{T}} G_1 \Pi_{\mathcal{B}}=0=\Pi_{\mathcal{B}} G_1 \Pi_{\mathcal{T}}.
\label{eq:S8_G1_block}
\end{equation}
Moreover, by construction each block generator commutes with its carrier,
\begin{equation}
[K_3,G_1|_{\mathcal{T}}]=0,\quad [K_2,G_1|_{\mathcal{B}}]=0,
\end{equation}
so that $[K,G_1]=0$ and $G_1$ is purely neutral. In the charge decomposition $G_1=\sum_q G_{1,q}$ this implies $G_{1,q\neq 0}=0$, and inserting into Eq.~\!\eqref{eq:S2_R} yields
\begin{equation}
R(\varepsilon)=\mathcal{O}(\varepsilon^2).
\end{equation}
Thus, for the mixed trimer-doublet protocol under the identical imperfection model, the leading charged corrections to both the trimer and doublet channels start at order $\varepsilon^2$.

\subsection{G2. Time-charge sectors and block-resolved observables}
\label{ssec:G2}

Let $\widetilde O=V(\varepsilon) O V^\dagger(\varepsilon)$ denote the dressed observable. Since the carrier $K$ is a direct sum across blocks and $D_0+\delta D(\varepsilon)$ is block diagonal in $(\Pi_{\mathcal T},\Pi_{\mathcal B})$, $\widetilde O$ decomposes as
\begin{equation}
\widetilde O=\widetilde O^{(\mathcal{T})}\oplus \widetilde O^{(\mathcal{B})},\qquad
\widetilde O^{(\mathcal{T})}=\Pi_{\mathcal{T}} \widetilde O \Pi_{\mathcal{T}},\qquad
\widetilde O^{(\mathcal{B})}=\Pi_{\mathcal{B}} \widetilde O \Pi_{\mathcal{B}},
\label{eq:S8_Osplit}
\end{equation}
up to corrections of order $\|R(\varepsilon)\|=\mathcal{O}(\varepsilon^2)$.

Within $\mathcal{T}$ we decompose $\widetilde O^{(\mathcal{T})}$ into $\mathbb{Z}_3$ time-charge sectors (Sec.~\!B1),
\begin{equation}
\widetilde O^{(\mathcal{T})}=\sum_{q=0}^{2}O^{(\mathcal{T})}_{q},\qquad
O^{(\mathcal{T})}_{q}=\frac{1}{3}\sum_{j=0}^{2}\omega^{-q j} K_3^{j}\widetilde O^{(\mathcal{T})}K_3^{-j},\qquad \omega=e^{2\pi i/3},
\label{eq:S8_grade3}
\end{equation}
so that $K_3 O^{(\mathcal{T})}_{q} K_3^{-1}=\omega^{q} O^{(\mathcal{T})}_{q}$. Similarly, within $\mathcal{B}$ we grade by $\mathbb{Z}_2$,
\begin{equation}
\widetilde O^{(\mathcal{B})}=\sum_{p=0}^{1}O^{(\mathcal{B})}_{p},\qquad
O^{(\mathcal{B})}_{p}=\frac{1}{2}\sum_{j=0}^{1}(-1)^{p j} K_2^{j} \widetilde O^{(\mathcal{B})} K_2^{-j},
\label{eq:S8_grade2}
\end{equation}
with $K_2 O^{(\mathcal{B})}_{p} K_2^{-1}=(-1)^{p} O^{(\mathcal{B})}_{p}$.

For the dressed magnetization $\widetilde M_z$ we use the shorthand
\begin{equation}
\widetilde M_z=\sum_{q=0}^{2}(M_z^{(\mathcal{T})})_{q}\ \oplus\ \sum_{p=0}^{1}(M_z^{(\mathcal{D})})_{p},
\label{eq:S8_Mzsplit}
\end{equation}
where $(M_z^{(T)})_{q}$ denotes the $q$th $\mathbb{Z}_3$ time-charge sector of the trimer contribution and $(M_z^{(D)})_{p}$ the $p$th $\mathbb{Z}_2$ sector of the doublet contribution. In the identical imperfection setting of Sec.~\!G1, the charged remainder satisfies $R(\varepsilon)=\mathcal{O}(\varepsilon^2)$, so to leading order these sectors generate independent $1/3$ and $1/2$ subharmonic components in the Fourier spectrum, with mutual mixing appearing only through subleading $\mathcal{O}(\varepsilon^2)$ corrections.

\subsection{G3. Period superposition in the dynamics}
\label{ssec:G3}
Let $\ket{\widetilde\psi_0}=V(\varepsilon)\ket{\psi_0}$ be the dressed initial state and decompose it as $\ket{\widetilde\psi_0}=\ket{\widetilde\psi_0^{(\mathcal{T})}}+\ket{\widetilde\psi_0^{(\mathcal{B})}}$ with weights $w_{\mathcal{T}}=\left\|\ket{\widetilde\psi_0^{(\mathcal{T})}}\right\|^2$ and $w_{\mathcal{B}}=\left\|\ket{\widetilde\psi_0^{(\mathcal{B})}}\right\|^2$. Using Eqs.~\!\eqref{eq:S8_nf}-\eqref{eq:S8_Mzsplit}, the stroboscopic expectation after $n$ periods splits additively,
\begin{equation}
\begin{aligned}
\langle M_z(n)\rangle
=&\sum_{q=0}^{2} e^{i2\pi q n/3} 
\left\langle \widetilde\psi_0^{(\mathcal{T})} \left| e^{i n D_{\mathcal{T}}} O^{(\mathcal{T})}_{q} e^{-i n D_{\mathcal{T}}} \right| \widetilde\psi_0^{(\mathcal{T})} \right\rangle
\\
+&\sum_{p=0}^{1} e^{i\pi p n} 
\left\langle \widetilde\psi_0^{(\mathcal{B})} \left| e^{i n D_{\mathcal{B}}} O^{(\mathcal{B})}_{p} e^{-i n D_{\mathcal{B}}} \right| \widetilde\psi_0^{(\mathcal{B})} \right\rangle
+\mathcal{O}\big(\|R(\varepsilon)\|\big),
\label{eq:S8_On}
\end{aligned}
\end{equation}
where $D_{\mathcal{T}}=\Pi_{\mathcal{T}} D_0 \Pi_{\mathcal{T}}$ and $D_{\mathcal{B}}=\Pi_{\mathcal{B}} D_0 \Pi_{\mathcal{B}}$. There are no cross terms between $\mathcal{T}$ and $\mathcal{B}$, because $\Pi_{\mathcal{T}} \Pi_{\mathcal{B}}=0$ and $[D_0,P]=0$. The leading oscillatory pieces are the charged sectors $q=1$ and $p=1$, which yield
\begin{equation}
\langle M_z(n)\rangle \approx 
A_{3}(n) e^{i2\pi n/3}
+A_{2}(n) e^{i\pi n}
+\text{c.c.}+E(n),
\label{eq:S8_mix}
\end{equation}
with slowly varying envelopes
\begin{equation}
A_{3}(n)=\left\langle \widetilde\psi_0^{(\mathcal{T})} \left| e^{i n D_{\mathcal{T}}} O^{(\mathcal{T})}_{1} e^{-i n D_{\mathcal{T}}} \right| \widetilde\psi_0^{(\mathcal{T})} \right\rangle,\quad
A_{2}(n)=\left\langle \widetilde\psi_0^{(\mathcal{B})} \left| e^{i n D_{\mathcal{B}}} O^{(\mathcal{B})}_{1} e^{-i n D_{\mathcal{B}}} \right| \widetilde\psi_0^{(\mathcal{B})} \right\rangle,
\label{eq:S8_A2A3}
\end{equation}
and a neutral part $E(n)$ sourced by $q=0$ and $p=0$. The relative contrast of the $1/3$ and $1/2$ lines tracks the block weights $w_{\mathcal{T}}$ and $w_{\mathcal{B}}$ together with the charged projections of $M_z$ in each block. For product preparations $\ket{\psi_0}=\bigotimes_i\ket{\phi_i}$ with fixed on-site weights in $\mathcal{T}$ and $\mathcal{B}$, $A_{3}(n)$ and $A_{2}(n)$ scale with the corresponding support and remain slowly varying over the locking window. This establishes the period superposition: if the initial state populates both blocks, the time trace contains concurrent $1/3$ and $1/2$ lines, and if it lies entirely in one block, only the corresponding subharmonic appears.

\section{G4. Noisy dynamics and experimental feasibility}
\label{sec:noisy_d5}

To evaluate the viability of our framework within state-of-the-art quantum architectures, we focus on the $d=5$ protocol with a system size of $N=10$ as the most phenomenologically rich realization of our model. This investigation moves beyond the ideal unitary regime to incorporate the dominant noise processes prevalent in contemporary superconducting-qudit platforms. We specifically examine the stability of the $1/d$ subharmonic response under the joint influence of coherent drive imperfections $\varepsilon$ and the accelerated decoherence characteristic of higher-level qudit states. By utilizing experimentally reported lifetimes and dephasing times as a computational benchmark, this analysis provides an exploratory assessment of the framework's potential robustness within realistic coherence windows.

In the noisy simulations, we mirror the experimental characterization of recent multilevel transmon qudits~\cite{Qudit_SC_LiuPRX2023,vbh4-lysv} by applying decoherence directly to the active energy levels. Specifically, we treat each elementary two-level rotation as an effective SU(2) operation and assign a localized quantum noise channel restricted to that driven subspace. This ensures our noise model accurately reflects how coherence is measured in current hardware. The channels account for the primary incoherent processes of energy relaxation and dephasing, which are modeled via amplitude damping governed by $T_1$ and pure-dephasing channels derived from the standard decomposition of $T_2$. To align our simulations with realistic hardware, we match the active subspaces of the $d=5$ protocol comprising $(0,2)$, $(1,3)$, and $(2,4)$ with reported coherence scales such as $(T_1,T_2) = (101, 37)~\mu\mathrm{s}$, $(73, 22.8)~\mu\mathrm{s}$, and $(73, 10)~\mu\mathrm{s}$. For the highest driven levels where data is sparse, we adopt values comparable to the nearest measured transitions as a representative estimate. Coupled with a gate duration of $t_{\mathrm{gate}}=0.12~\mu\mathrm{s}$, this construction establishes an experimentally grounded mapping from measured coherence properties to the gate-level noise strengths in our analysis~\cite{Qudit_SC_LiuPRX2023,vbh4-lysv}.

\begin{figure}[htbp]
    \centering
    \includegraphics[width=0.75\linewidth]{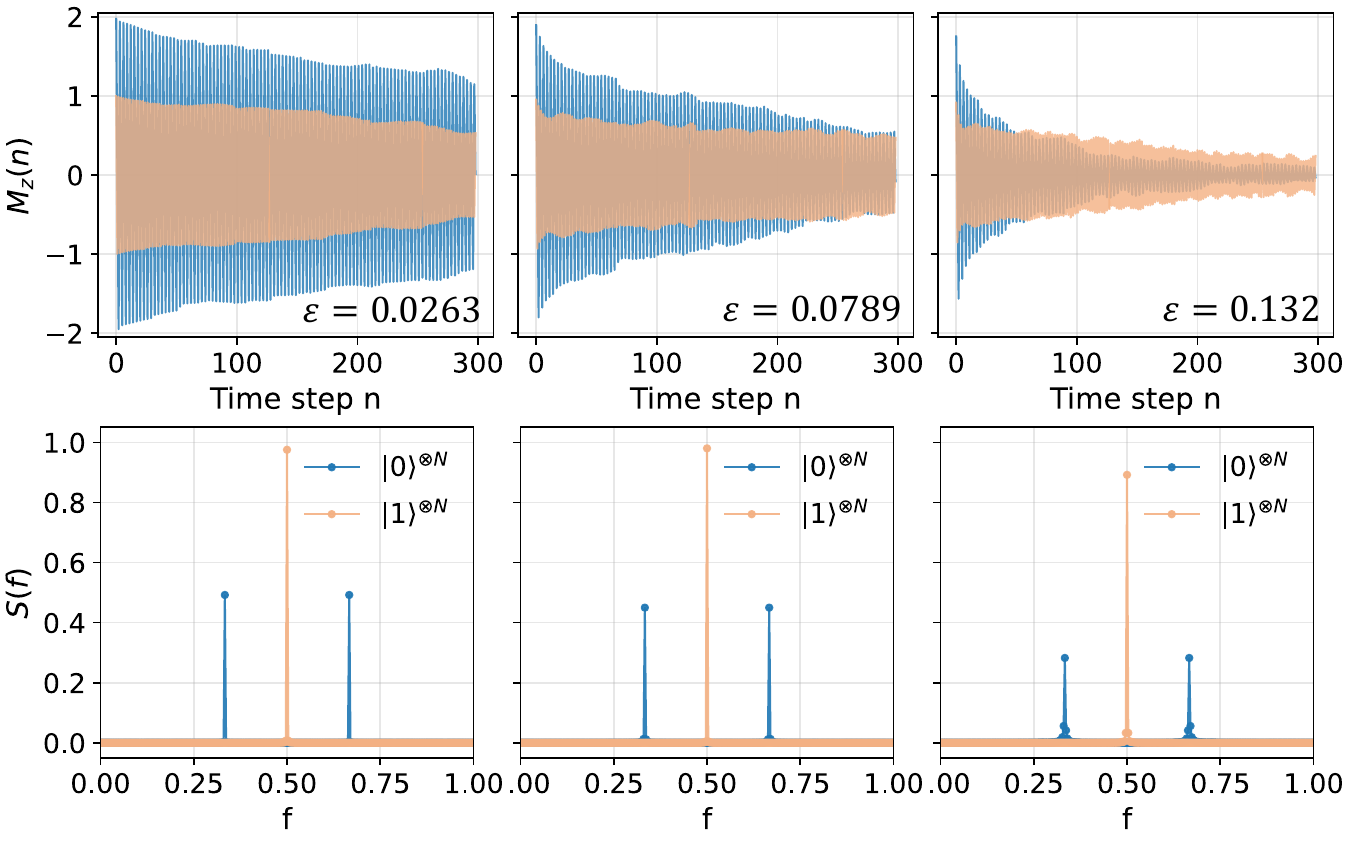}
    \caption{Numerical simulation of the $d=5$ protocol for $N=10$ sites under realistic noise profiles, with relaxation and dephasing applied directly to the active energy levels. The analysis utilizes gate-level noise strengths mapped from reported coherence scales in multilevel superconducting platforms, where the driven subspaces $(0,2)$, $(1,3)$, and $(2,4)$ are assigned $(T_1, T_2)$ values of $(101, 37)~\mu\mathrm{s}$, $(73, 22.8)~\mu\mathrm{s}$, and $(73, 10)~\mu\mathrm{s}$, respectively. At a gate duration of $t_{\mathrm{gate}} = 0.12~\mu\mathrm{s}$, the stroboscopic magnetization $M_z(n)$ exhibits robust subharmonic oscillations that persist over multiple Floquet periods before undergoing gradual decay. The corresponding spectral response retains identifiable peaks near the subharmonic frequency even as increased coherent drive imperfections $\varepsilon$ accelerate the signal attenuation.}
    \label{fig:noisy_d5}
\end{figure}

The resulting dynamics are shown in Fig.~\!\ref{fig:noisy_d5}. Even in the presence of the above incoherent processes, the system continues to exhibit a clear subharmonic response over the time window. In the time domain, the stroboscopic magnetization $M_z(n)$ displays robust oscillations persisting for periods before gradually decaying, rather than collapsing immediately once decoherence is introduced. In the frequency domain, the spectral response retains pronounced peaks near the subharmonic frequency, indicating that the period-doubled structure remains clearly identifiable on experimentally relevant timescales. As the coherent drive imperfection $\varepsilon$ is increased, the oscillation envelope decreases more rapidly, but the characteristic frequency structure remains visible in the regime shown in Fig.~\!\ref{fig:noisy_d5}. These results show that, even in the presence of experimentally motivated relaxation and dephasing, the subharmonic locking remains observable over a finite time window.

In realistic devices, secondary effects such as residual long-range couplings, frequency crowding, off-resonant driving, and control cross-talk can further degrade signal coherence and broaden spectral features. These mechanisms are well-documented in transmon-based qudit experiments~\cite{Qudit_SC_LiuPRX2023,vbh4-lysv}. However, even with the availability of modeling techniques~\cite{mapra2024}, a high-fidelity replication of physical platforms remains challenging due to the evolving nature of multi-level noise characterization and the scarcity of device-specific parameters. A full calibration-level description of such imperfections, therefore, exceeds the scope of the present theoretical study. Nevertheless, by incorporating the most fundamental limitation of reduced coherence times, our noisy simulations provide an estimate of the experimentally accessible regime. In this sense, our results provide a necessary bridge between idealized theoretical models and the constraints of physical hardware, confirming that the $d=5$ protocol remains observable on near-term multilevel superconducting-qudit platforms, even if a fully optimized realization requires further engineering considerations.

\providecommand{\noopsort}[1]{}\providecommand{\singleletter}[1]{#1}%